\documentclass[10pt,twocolumn,aps,prl,superscriptaddress,showpacs,tightenlines,pdflatex,longbibliography]{revtex4-2}
\usepackage{upgreek}
\usepackage{amsthm}
\usepackage{gensymb}
\usepackage{braket}
\usepackage{amsmath}            
\usepackage{amssymb}           
\usepackage{mathtools}
\usepackage{graphicx}           
\usepackage[caption=false]{subfig}
\usepackage{xcolor}
\usepackage{tcolorbox}
\usepackage{enumitem}
\usepackage[T1]{fontenc}

\usepackage{etoolbox}

\usepackage{enumitem}
\usepackage{xcolor}
\usepackage{bm}
\usepackage{times}
\usepackage{newtxmath}

\makeatletter
\let\cat@comma@active\@empty

\usepackage{hyperref} %% For arXiv

\newcommand{\id}{\openone}

\newcommand{\td}[1]{\widetilde{#1}}
\newcommand{\II}{\mathcal{I}}
\newcommand{\vA}{\vec{A}}
\newcommand{\mean}[1]{\langle #1 \rangle}
\newcommand{\SR}{\mathbb{SR}}

\newcommand{\IR}{\mathbb{IR}}
\newcommand{\JM}{\mathbb{JM}}
\newcommand{\LHS}{\mathbb{LHS}}
\newcommand{\ROM}{\mathbb{ROM}}

\newcommand{\SW}{\mathbb{SW}}

\newcommand{\IW}{\mathbb{IW}}
\DeclareMathOperator{\tr}{tr}
\newcommand{\LFo}{{\rm LF_1}}
\newcommand{\LF}{{\rm LF}}
\newcommand{\EE}{\mathcal{E}}
\newcommand{\Ss}{\mathcal{S}}
\newcommand{\ketbrac}[1]{|#1\rangle\langle #1|}
\newcommand{\ketbra}[2]{\left \vert #1 \right \rangle \hspace{-0.4em} \left \langle #2 \right \vert}

\newtheorem{theorem}{Theorem}

\newtheorem{lemma}[theorem]{Lemma}
\newtheorem{result}{Result}
\newtheorem{obs}{Observation}

\newcommand{\costa}[1]{{\color{magenta}#1}}

\begin{document}
\title{Measurement incompatibility signature cannot be stochastically distilled}

\author{Huan-Yu Ku}
\affiliation{
Department of Physics, National Taiwan Normal University, Taipei 11677, Taiwan}

\author{Chung-Yun Hsieh}
\email{chung-yun.hsieh@bristol.ac.uk}
\affiliation{H.H. Wills Physics Laboratory, University of Bristol, Tyndall Avenue, Bristol BS8 1TL, United Kingdom}

\author{Costantino Budroni}
%\email{costantino.budroni@unipi.it}
\affiliation{Department of Physics ``E. Fermi'' University of Pisa, Largo B. Pontecorvo 3, 56127 Pisa, Italy}

\begin{abstract}
We show that the incompatibility of a set of measurements cannot be increased by subjecting them to a filter, 
namely, by combining them with a device that post-selects the incoming states on a fixed outcome of a stochastic 
transformation.
This result holds for several measures of incompatibility, such as those based on robustness and convex weight. 
Expanding these ideas to Einstein-Podolsky-Rosen steering experiments, we are able to solve the problem of the maximum steerability obtained with respect to the most general local filters in a way that allows for an explicit calculation of the filter operation. Moreover, our results generalize to nonphysical maps, i.e., positive but not completely positive linear maps.

\end{abstract}
\maketitle

\date{\today}
%{\it Introduction.---} 
\section{Introduction}
A remarkable aspect of quantum mechanics is that certain physical quantities, e.g., position and momentum of a particle, cannot be simultaneously measured: They are said to be incompatible.  
Measurement incompatibility is at the core of many quantum properties, from uncertainty relations~\cite{Busch2014RMP,Otfried2021Rev}, to Bell nonlocality~\cite{Wolf2009PRL,Brunner2014RMP,QuintinoPRL2019,Shin-Liang2021PRR}, Kochen-Specker contextuality~\cite{BudroniRMP2022}, Einstein-Podolsky-Rosen steering~\cite{Uola2014PRL,Quintino2014PRL,Uola2015PRL,Zhao2020,Lee2023arXiv}, quantum complementarity~\cite{Hsieh2023-3}, and so on. 
Importantly, it is an indispensable resource in many operational tasks in quantum science and technologies, e.g., cryptography~\cite{Bennett92,Hsieh2023-3}, state discrimination~\cite{BarnettAOP09,Bae2011PRL,Bae_2015Review,Bae2019PRL,Paul2019PRL,Carmeli2019PRL,Uola2019PRL}, measurement simulation~\cite{Buscemi2020PRL}, and steering distillation~\cite{Ku2022NC,Hsieh2023,Wang2023arXiv}.
Thus, understanding how to harness this crucial resource is essential.
In particular, it is of great practical value to know {\em how to operationally distill incompatibility}. 
That is, when our physical devices are only weakly incompatible, can we use experimentally allowed operations to enhance incompatibility?
A clear answer can identify the cost of preparing incompatibility as well as uncover the fundamental limit on manipulating incompatibility. 

Even if incompatibility distillation has never been studied, theorists have investigated how to manipulate incompatibility via physically motivated and practically implementable operations ~\cite{PuseyJOPSA15,Heinosaari2015PRA,Haapasalo2015,Carmeli2019PRL,
DesignolleNJP2019,Uola2019PRL,Paul2019PRL,Buscemi2020PRL,Uola2020PRL, Ducuara2020PRL}. 
These include quantum pre-processing \cite{Heinosaari2015PRA}, conditional classical pre- and post-processing~\cite{Paul2019PRL}, and their combination~\cite{Buscemi2020PRL}. 
These findings all suggest that incompatibility of a {\em single-copy} system cannot be increased {\em deterministically} by a broad range of allowed operations.
Hence, crucially, to distill incompatibility, the allowed operations need to be either {\em multi-copy} (increasing the physical system size) or {\em stochastic} (increasing the waiting time in the experiment as the protocols become probabilistic).
While the former is usually practically challenging and expensive~\cite{Peres1996PRA,Bennett1996,Pan2001,Pan2003}, stochastic distillation by filtering operation has been recognized as an experimentally feasible method to enhance various quantum resources, e.g., entanglement~\cite{Horodecki1999,Wang2006PRL,Masanes2008PRL,GUHNE20091,Kim2011,Thinh2018PRA,Ku2021PRR}, nonlocality~\cite{Kwiat2001,Wang:20}, steering~\cite{PramanikPRA2019,Nery2020,Zhang2023arXiv,Hao2024PRA}, and teleportation power~\cite{Horodecki2000PRA,Li2021}.
It is thus vital to know whether measurement incompatibility can be distilled by filtering operations, which are stochastic operations followed by a post-selection on a fixed outcome.

In this work, we show that, in stark contrast to most of the quantum resources, filters {\em cannot increase} measurement incompatibility under all known canonical figure-of-merits, i.e., robustness and weight-based measures~\cite{Paul2019PRL,Carmeli2019PRL,Uola2019PRL,Heinosaari2015PRA,Haapasalo2015, DesignolleNJP2019,PuseyJOPSA15, Uola2020PRL, Ducuara2020PRL}.
%\HY{To generalize our statement, this result is also held in a single measurement scenario~\cite{Paul2019PRL2}.}
Thus, these serve as a strong no-go result, as it suggests that there is arguably no physically cheap way to distill incompatibility --- one either needs to involve multi-copy settings (experimentally expensive) or more general mathematical operations/figure-of-merits (that can easily lack physical meanings and/or be practically challenging).

We extend this investigation to steering~\cite{Wiseman2007PRL,UolaRMP2020,Cavalcanti2016,Xiang2022PRXQuantum}, where our results solve the problem of the maximum steerability obtainable under the {\em most general local filters}.
In particular, we show that a filter with the single Kraus operator is sufficient to solve the problem, namely increasing the number of Kraus operators cannot improve stochastic steering distillation.
Consequently, we can compute the optimal filter explicitly from the results of Ref.~\cite{Ku2022NC}.
Finally, our results also hold for {\em nonphysical} filters, i.e., positive linear maps that are not completely positive.

%{\it Preliminary notions.---}
\section{Preliminary notions}

In quantum mechanics, a measurement is a procedure that generates a classical outcome with a certain probability for any input state. In abstract terms, we can denote a measurement procedure as a mapping from quantum states to probability distributions, namely,
\begin{equation}\label{eq:meas_BR}
\mathcal{M} : \rho \mapsto \{ p(a) \}_a.
\end{equation}
Clearly, $\{p(a)\}_a$ satisfies $p(a)\geq 0$ for all $a$ and $\sum_a p(a)=1$. Moreover, this mapping must be linear \footnote{In reality, the constraint implies that it is convex linear, however, we can always extend it by linearity, since the extension involves objects that are not state, so they do not have a physical interpretation.}, since if I consider a probabilistic mixtures of states $\rho=\lambda \rho_1 + (1-\lambda) \rho_2$ with $0 < \lambda <1$, it must be $p(a)=\lambda p_1(a)+(1-\lambda)p_2(a)$, and bounded. This implies that it can be written as a scalar product \footnote{By Frechét-Riesz theorem, any continuous linear functional $\varphi$ on a Hilbert space is represented by a vector $\phi$ through the scalar product $\varphi(\psi)=\mean{\phi|\psi}$. In this case, the scalar product on the space of linear operators is given by  $\mean{A,B}=\tr[A^\dagger B]$.}, namely,
\begin{equation}\label{eq:Born_rule}
p(a)=\tr[M_a \rho],
\end{equation} 
which is commonly known as the {\it Born rule}. 
To satisfy the positivity and normalization constraints for $\{p(a)\}_a$, it must be $M_a \geq 0$ for all $a$ and $\sum_a M_a=\openone$. These conditions define a {\it positive operator-valued measure} (POVM). For more details on this construction, see, e.g., \cite{Heinosaari-Ziman} Ch. 2. The same axiomatic derivation is at the basis of several generalizations of the idea of a quantum measurement, namely, the formulation of higher-order operations~\cite{ChiribellaEPL2008}; for instance, see the construction of process matrices in \cite{AraujoNJP2015}.

Note that, in this construction and in particular in the derivation of the Born rule in Eq.~\eqref{eq:Born_rule}, a crucial role is played by the linearity of the mapping $\rho \mapsto p(a)$. When the mapping is not linear, as it is the case for experiment with postselection, e.g., filtering, the Born rule and, consequently, the representation of a measurement as a POVM is no longer guarantee. We will come back to this point.

In the general case of multiple measurements, a fundamental notion is that of {\it measurement incompatibility}. The incompatibility of quantum measurements refers to the impossibility of accessing experimental outcomes simultaneously e.g., the position and momentum of a quantum system. Considering multiple inputs $x$, so we have a collection of POVMs $\vec{M}:=\{M_{a|x}\}_{a,x}$, which we call a {\it measurement assemblage} (MA). If there exists a single POVM $\{G_\lambda\}_{\lambda}$ that can fully characterize the set of POVMs by classical post-processing, the MA is said to be jointly measurable (JM) i.e., $M_{a|x}=\sum_\lambda p(a|x,\lambda)G_\lambda$, where $p(a|x,\lambda)$ is a conditional probability. If there is no such description, the measurement assemblage is said to be {\it incompatible}. Incompatibility can be quantified in different ways, for instance via the
{\it incompatibility robustness}~\cite{Uola2015PRL,Haapasalo2015} defined as 
\begin{equation}
\begin{aligned}
\IR(\vec{M})&\coloneqq\min\{t \geq 0\;|\;\exists \vec{N}~\text{MA and}~\vec{D}\in\mathbb{JM}\ \text{s.t.}\\
&\ (M_{a|x} + t N_{a|x})/(1+t)= D_{a|x}\ \forall~a,x\}, 
\end{aligned}
\end{equation}
where $\mathbb{JM}$ denotes the set of JM measurement assemblages. Several other measures can be introduced by putting restrictions on the MA $\vec{N}$; see, e.g., Ref.~\cite{DesignolleNJP2019}, or using different convex decompositions~\cite{PuseyJOPSA15}.

Steering is a protocol in which one party, Alice, performs measurements $\vec{A}=\{A_{a|x}\}_{a,x}$ on her half 
of a bipartite system in the initial state $\rho_{\rm AB}$ and sends the measurement information, i.e., input $x$ 
and output $a$, to Bob. In this way, Bob obtains a collection of local states labeled as 
$\vec{\sigma}:=\{\sigma_{a|x}\}_{a,x}$, where $\sigma_{a|x}=\tr_{\rm A}[\rho_{\rm AB} (A_{a|x}\otimes \openone)]$ is in an unnormalized form. We call $\vec{\sigma}$ a state assemblage (SA). It satisfies 
$\sigma_{a|x}\geq 0, \forall a,x$ and $\sum_{a} \sigma_{a|x}=\rho_{\rm B}=\tr_A[\rho_{\rm AB}], \forall x$. A SA 
is {\it unsteerable} if there is a {\it local-hidden-state} (LHS) model, namely, if we can write 
$\sigma_{a|x}=\sum_\lambda p(\lambda)p(a|x,\lambda)\rho_{\lambda}, \forall a,x$, and {\it steerable} otherwise. 
Steering can be quantified, for instance, by the {\it steering robustness}~\cite{Piani2015} defined as 
\begin{equation}
\begin{aligned}
\SR(\vec{\sigma})&\coloneqq\min\{t\geq 0\;|\; \exists\; \vec{\xi}\; \text{SA and},\; \vec{\tau}\in\LHS~\text{s.t.}\\&~(\sigma_{a|x} + t \xi_{a|x})/(1+t)=\tau_{a|x}\ \forall~ a,x\}
\end{aligned}
\end{equation}
where $\mathbb{LHS}$ is the set of SA admitting a LHS model.

The SA $\vec{\sigma}$ is unsteerable, whenever $\rho_{\rm AB}$ is separable or $\vec{A}$ is JM, i.e., both entanglement and measurement incompatibility are necessary for steering. This connection is indeed much deeper \cite{Uola2014PRL,Quintino2014PRL} and steering and incompatibility can be seen as equivalent problems \cite{Uola2015PRL}. A central notion in this correspondence is that of {\it steering-equivalent-observables} (SEOs)~\cite{Uola2015PRL} which map a SA to a MA as
\begin{equation}
B_{a|x} := \rho_{\rm B}^{{-1/2}} \sigma_{a|x} \rho_{\rm B}^{{-1/2}}, \text{ for } \rho_{\rm B} \text{ full-rank,}
\end{equation}
or an analogous expression with an additional isometry in case $\rho_{\rm B}$ is not full rank; see Ref. \cite{Uola2015PRL} for more details. It is straightforward to see that SEOs define equivalence classes among SAs, i.e., via the relation of having the same SEO. These equivalence classes were recently studied from the perspective of local filters \cite{Ku2022NC}. 
We recall that a filter corresponds, operationally, to performing a measurement and post-selecting on a specific outcome. 
Mathematically, this operation is represented by a completely positive (CP) and trace-nonincreasing (TNI) map, $\EE_{\omega}$, followed by a post-selection of the state: 
\begin{equation}\label{Eq: state transform after LF}
\rho \mapsto \frac{\EE_{\omega}(\rho)}{\tr[\EE_{\omega}(\rho)]}.
\end{equation}
Crucially, the filter's action on quantum states must include a post-selection, which leads to a nonlinear transformation. This is unavoidable and it plays a central role at the moment of defining a notion of {\it filtered measuremet}, as we discuss below. 
Physically, $\EE_{\omega}$ can be viewed as part of a bigger measurement $\{\EE_{\omega}\}_\omega$, which is a {\em quantum instrument} (each $\EE_\omega$ is CP and $\sum_\omega \EE_\omega$ is CPTP), and we use $\omega$ to specify the measurement outcome.
Importantly, it was shown in Ref.~\cite{Ku2022NC} that such classes also define which SAs can be transformed into each other via {\em $\LFo$ operations}, which are filters of the form $\EE_{\omega}(\cdot) = K(\cdot)K^\dagger$ with $K^\dagger K\le\id$, i.e., local filters consisting of only one Kraus operator.
Moreover, the maximal steerability within one class corresponds to the incompatibility of the SEO, namely,
\begin{equation}\label{eq:max_SR=IR}
\max_{\EE_\omega \in \LFo}\SR\left(\frac{\EE_\omega(\vec{\sigma})}{p(\omega)}\right)=\max_{\rho_{\rm B}} \SR\left(\rho_{\rm B}^{{1/2}} \vec{B} \rho_{\rm B}^{{1/2}}\right) = \IR(\vec{B}),
\end{equation}
where $\vec{B}$ is the SEO of $\vec{\sigma}$, $\omega$ represents the outcome on which one post-selects with probability \mbox{$p(\omega)=\tr[\EE_\omega( \sum_a \sigma_{a|x} )]$;} see Appendix A. 
In the following, when we talk about steerability (incompatibility) we implicitly refer to the steerability (incompatibility) robustness, unless explicitly stated otherwise.
This is indeed the most widely used steering measure, from which experimentally testable steering inequalities can be derived ~\cite{Cavalcanti2016,Hsieh2016,Ku2018PRA,Tendick2023distancebased}. We thus focus on it first, but our results generalize to other measures, as we discuss later on.

%%%%%%%%%%%%%%%%%%%%%%%%%%%%%%%%%%%%%%%%%%%%%%%%%%%%%%%%%%%%%%%%%%%%%%%%%%%%%%%%
\begin{figure}[t]
\includegraphics[width=1\columnwidth]{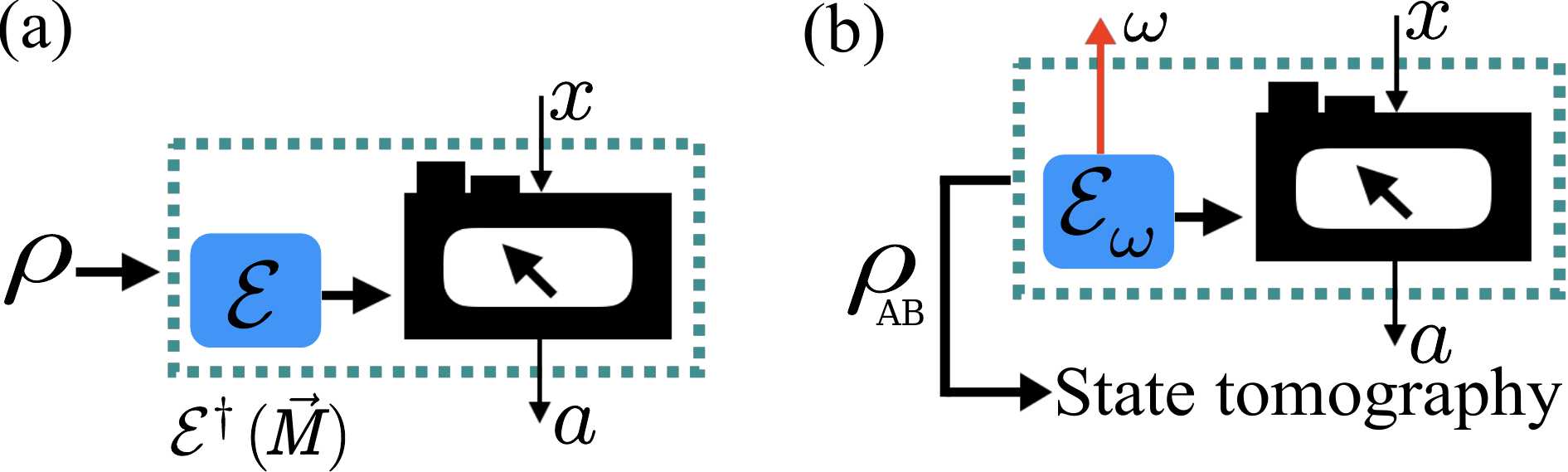}
\caption{(a) Channel applied before a measurement. The channel can be seen as applied to the state (Schr\"odinger picture) or applied to the measurement (Heisenberg picture). (b) Analogous setup, but with an instrument. The instrument can still be seen equivalently as part of the state preparation or the measurement procedure. 
The measurement incompatibility is operationally defined by considering the filtered measurement in a steering protocol.} 
\label{Fig1}
\end{figure}
%%%%%%%%%%%%%%%%%%%%%%%%%%%%%%%%%%%%%%%%%%%%%%%%%%%%%%%%%%%%%%%%%%%%%%%%%%%%%%%%

%{\em 
% Operational interpretation and quantification of the filtering of measurements.---
%Filtering measurements and operationally quantifying their incompatibility.---
%}
\section{Filtering measurements and operationally quantifying their incompatibility}
The first step to derive our result is to reason what it means to {\em filter} a measurement.
In the Schr\"odinger picture, filtering a state $\rho$ corresponds to the application of a CPTNI map $\EE_\omega$, as $\rho\mapsto\EE_{\omega}(\rho)/\text{tr}[\EE_{\omega}(\rho)]$. We emphasize, again, that the non-linear transformation is necessary. In contrast in the Heisenberg picture, the direct application of the adjoint map on a POVM $\vec{M}$, i.e., $\{\EE^\dagger_{\omega}(M_a)\}_a$, is not possible, as this provides an object that is not properly normalized $\sum_a \EE^\dagger_{\omega}(M_a)\neq \openone$. 
In order to derive the correct mathematical representation for the filtered measurement, one has to start from the physical procedure of a measurement in quantum mechanics.
In the lab, we perform the filter between the state preparation $\rho$ and the measurement $\{M_a\}_a$. When the correct outcome $\omega$ is not observed, we discard all experimental rounds.
The analog of Eq.~\eqref{eq:meas_BR} for the case of both filter and measurement, then, can be written as 
\begin{equation}\label{beyondPOVM}
\rho \mapsto \{ p(a) \}_a=  \left\lbrace \frac{\text{tr}[M_a\EE_{\omega}(\rho)]}{\text{tr}[\EE_{\omega}(\rho)]} \right\rbrace_a = \left\lbrace\frac{\tr[\EE^\dagger_\omega(M_a)\rho]}{\tr[\EE^\dagger_\omega(\openone)\rho]} \right\rbrace_a,
\end{equation}
associating to arbitrary state $\rho$ a probability $p(a)$.
This expression can be seen as the generation of the Born's rule when post-selection is involved, since when $\mathcal{E}_\omega$ is CPTP, it returns to the standard Born's rule. Note that Eq.~\eqref{beyondPOVM} gives a unique and state-independent definition of filtered measurement.
The nonlinearity of the map is unavoidable, as the measurement procedure involves a post-selection on the filter being passed. In particular, this implies that the mapping between states and probabilities induced by a filtered measurement can no longer be mathematically described by linear operators (POVMs) in the trace expression, as in Eq.~\eqref{eq:Born_rule} (see also the explicit discussion of non-POVM property of a filtered measurement by Arai and Hayashi~\cite{Hayashi2024arXiv}).

A natural question is, then, how to define incompatibility for filtered measurements, if they can no longer be described by POVMs. For this, reasoning in terms of operations in the lab turns out to be crucial. More precisely, we can reason in terms of quantum information tasks for which incompatibility is a resource and ask the question: \emph{can we improve the performances of our measurement devices by applying a filter?} An affirmative answer denotes an enhancement of the resource, i.e., measurement incompatibility. One of such tasks is quantum steering \cite{Uola2014PRL,Quintino2014PRL,Uola2015PRL}. In particular, it has been shown \cite{Ku2022NC} that, for a given MA $\vec{A}$, $\IR$ can be operationally characterized in a steering scenario as:
\begin{equation}\label{eq:IR_sup_SR}
\IR(\vec{A}) = \max_{\rho_{\rm AB}}\ \SR(\vec{\sigma}(\rho_{\rm AB})),
\end{equation}
with $\{\sigma_{a|x}(\rho_{\rm AB})\}_{a,x}$ the SA obtained by measuring $\vec{A}$ on the bipartite state $\rho_{\rm AB}$; see Appendix A for details. In simple terms, the incompatibility of measurements is understood as the maximal steerability that can be obtained when applying such measurement to a bipartite quantum state. With such an operational characterization, not directly involving the notion of POVM for the filtered measurement, we can proceed to prove our main result.

%{\it Measurement incompatibility cannot be stochastically distilled.---}
\section{Measurement incompatibility cannot be stochastically distilled}
According to Eq.~\eqref{eq:IR_sup_SR}, incompatibility is the maximum steerability over all bipartite states. Then, applying the definition 
of filtered measurement, i.e., Eq.~\eqref{beyondPOVM}, to the steering scenario, we transform the SA
\begin{equation}
\sigma_{a|x}(\rho_{\rm AB})\mapsto \sigma_{a|x}(\rho'_{\rm AB}) =\frac{\tr_{\rm A}[(A_{a|x}\otimes \openone)(\EE_\omega\otimes\openone)\rho_{\rm AB}]}{{\rm tr}[(\EE_\omega\otimes\openone)\rho_{\rm AB}]}
\end{equation}

The effect of the filter is then to modify the initial state $\rho_{\rm AB}\mapsto \rho'_{\rm AB}$. Since we are considering $\max_{\rho_{\rm AB}}$, this can never increase the value of the right-hand side of Eq~\eqref{eq:IR_sup_SR}.
We thus obtained
\begin{result}\label{R:inc_dist}
Measurement incompatibility cannot be increased by filters;
i.e., it cannot be stochastically distilled.
\end{result}

This result requires a recently proposed operational approach to quantify measurement incompatibility \cite{Ku2022NC}: measurement incompatibility is equivalent to optimal steering distillation under local filters. 
Notice that, even if we use a steering scenario to describe and quantify measurement incompatibility operationally, the no-go Result~\ref{R:inc_dist} is independent of the specific task chosen. For instance, the same argument applies when defining
incompatibility robustness in terms of state discrimination tasks. This evolves a more complicated structure due to the nonlinearity in the definition~\cite{Paul2019PRL,Buscemi2020PRL}; see Appendix C.

Moreover, analogous arguments apply to different scenarios. First, one can show that the incompatible weight~\cite{PuseyJOPSA15}, which has an operational meaning in both quantum steering and state-exclusion tasks~\cite{Uola2020PRL,Ducuara2020PRL}, cannot be distilled by filtering.
Second, a similar no-go result hold when there is a single measurement involved, i.e., independently of incompatibility. we show that our argument is also valid for {\it an informativeness measurement} (a property of a \emph{single} measurement). This is the case of the {\it robustness of measurement} \cite{Paul2019PRL2}, i.e., the minimal amount of noise that needs to be added to the given single measurement before it becomes trivial and can be operationally defined via state-discrimination tasks. Also in this case, we show that it cannot be increased by filtering (see Appendix C).
We thus uncover a series of general no-go results for measurement resources under filters. 

Our results imply that once $\EE_\omega$ is a positive linear map, it cannot distill incompatibility.
Hence, at a mathematical level, our no-go result also rules out various non-physical linear maps, such as positive linear maps that are not CP or trace-increasing
(in the latter case, the physical interpretation is more problematic since one cannot make sense of the notion of post-selection and the associated renormalization of probability).
Finally, our result provides a clear physical insight: multi-copy measurement settings are essential for distilling measurement resources. In fact, the use of multiple copies of a measurement device has also been recently discussed in relation to simulating a target measurement using multiple imperfect copies~\cite{Paul2022, Joonwoo2024arXiv,Corlett2024arXiv}.

%{\it Filtering operations on quantum steering.---}
\section{Filtering operations on quantum steering}
Here, we discuss the effect of positive linear maps, which we later on interpret as filters, on the different elements of a SA. The goal is to show that local filters described by a single Kraus operator, denoted by $\LFo$, are always enough to obtain the maximum steerability over all filter operations, denoted by $\LF$.
We start with the following simple observation:
\begin{obs}\label{obs:marginal are the same}
For a reduced state $\rho_{\rm B}$ such that $\SR(\rho_{\rm B}^{{1/2}} \vec{B} \rho_{\rm B}^{{1/2}}) = \IR(\vec{B})$, as in Eq.~\eqref{eq:max_SR=IR}, we can choose $\vec{\xi}$ and $\vec{\tau}$ appearing in the definition of $\SR(\rho_{\rm B}^{{1/2}} \vec{B} \rho_{\rm B}^{{1/2}})$ %coincide with
such that their reduced states coincide with $\rho_{\rm B}$, i.e., $\sum_{a} \xi_{a|x}=\sum_a \tau_{a|x}=\rho_{\rm B}$. 
\end{obs}
We detailed the proof in Appendix D. 
This observation suggests that the definitions of $\IR$ and $\SR$ are completely equivalent when we consider the maximal element in an SEO class. Moreover, this observation implies that the {\it consistent steering robustness} $\SR^{\rm c}(\vec{\sigma})$~\cite{Cavalcanti2016PRA} (a robustness with the requirement that the noise term  $\vec{\xi}$ has the same reduced state as the original SA) quantifies the maximal steerability over the SEO class of $\vec{\sigma}$, or equivalently, over all possible SA that can be reached via local filters. Indeed, it has been shown that $\SR^{\rm c}(\vec{\sigma})=\IR(\vec{B})$, see \cite{ShinLiang2016PRL,ShinLiang2018PRA}.

From Obs.~\ref{obs:marginal are the same}, another useful result follows:

\begin{result}\label{R:mathfilt}
Given an MA $\vec{B}$ and a positive map $\EE_{\omega}$, we have
\begin{align}
\begin{split}\label{Eq:E_dag}
\max_{\rho_{\rm{B}}}\SR\left(\frac{\rho_{\rm{B}}^{1/2}\EE_{\omega}^\dagger(\vec{B})\rho_{\rm{B}}^{1/2}}{p(\omega)}\right) \leq \max_{\rho_{\rm{B}}} \SR\left(\rho_{\rm{B}}^{1/2}\vec{B}\rho_{\rm{B}}^{1/2}\right),
\end{split}\\
\begin{split}\label{Eq:E}
\max_{\rho_{\rm{B}}}\SR\left(\frac{\EE_{\omega}(\rho_{\rm{B}}^{1/2}\vec{B}\rho_{\rm{B}}^{1/2})}{p(\omega)}\right) \leq \max_{\rho_{\rm{B}}} \SR\left(\rho_{\rm{B}}^{1/2}\vec{B}\rho_{\rm{B}}^{1/2}\right),
\end{split}
\end{align}
where $p(\omega)=\tr[\EE_\omega(\rho_{\rm{B}})]$.
\end{result}
See Appendix E for the proof.
So far, this result remains at some abstract level. For instance, notice that $\EE_{\omega}$ is not necessarily a physical map, i.e., CP and trace-non-increasing, but just a positive linear map.
To see its physical significance, (in Appendix F) we show that $\LFo$ are enough to obtain maximum steerability:
\begin{result}\label{R:LF1 is optimal}
The maximal steerability obtainable by $\LFo$ filtering, is the same as the one obtainable with general filters in $\LF$, namely,
\begin{equation}\label{eq:LF1=LF}
\max_{\EE_\omega \in \LFo}\SR\left(\frac{\EE_\omega(\vec{\sigma})}{p(\omega)}\right) = \max_{\EE_\omega \in \LF}\SR\left(\frac{\EE_\omega(\vec{\sigma})}{p(\omega)}\right).
\end{equation}
\end{result}

This result completes the classification of steerability with respect to local operations presented in Ref.~\cite{Ku2022NC}: the maximal steerability obtainable with $\LFo$ filters 
is the same as the one obtainable with general $\LF$ filters.
This completely solves the problem of finding optimal local filters for stochastic steering distillation, since the optimal LF1 can be efficiently computed by a semi-definite program (SDP)~\cite{Ku2022NC,CVXBook,SDP-textbook}. 
Notice that $\LF$ filters are able to change the SEO class of the SA, contrary to what happens with two-way $\LFo$ \cite{Ku2022NC}, and that Result~\ref{R:LF1 is optimal} holds for any positive linear map $\EE_\omega$. 
Finally, it seems that Res.~\ref{R:LF1 is optimal} can be obtained by decomposing $\LF$ filters individually and use the optimal one to achieve left-hand side of Eq.~\eqref{eq:LF1=LF}. We emphasize it is not the case due to the non-convexity and non-linearity of the transformation. Hence, the decomposition of $\LF$ cannot be directly applied to the $\SR$.

The second physical implication of Result~\ref{R:mathfilt} reads
\begin{result}\label{R:steer_filt}
Consider an experiment where Alice steers Bob's state
via a pure bipartite state $\rho_{\rm AB}=\ketbrac{\psi}_{\rm AB}$. 
If Bob's assemblage $\vec{\sigma}$ is already optimal w.r.t.~Bob's local operations, i.e., $\SR(\vec{\sigma})=\max_{\EE_\omega \in \LF}\SR\left(\frac{\EE_\omega(\vec{\sigma})}{p(\omega)}\right)$, then its steerability cannot be increased by performing any filter on Alice's side.
\end{result}
See Appendix G for the proof.
This result states that for a bipartite pure state and whenever steerability is already maximized by performing some local filter on Bob's side, Alice cannot further increase it by performing local filters. 
This result can also be seen as the impossibility for Alice to distill the incompatibility of the MA using local filters, aligning with the findings in Res.~\ref{R:inc_dist}. 
Notice that this is no longer true if Alice and Bob share a mixed state and 
indeed the steerability can be increased via local filters on Alice's side, as it is the case for the 
phenomenon of {\it hidden steerability} \cite{QuintinoPRA2015,PramanikPRA2019}. Indeed, trying to argue 
in terms of a purification of the initial state does not work. This phenomenon is not restricted to 
quantum steering, but it was also observed in Bell 
nonlocality~\cite{Popescu1995PRL,Gisin1996PLA,Liang2012,Hirsch2013PRL} and quantum 
teleportation~\cite{Li2021}.

%{\it Extension to other incompatibility measures.---}
\section{Extension to other incompatibility measures}
Our results are not limited to the (incompatibility/steering) robustness, but they apply to several other measures.
We start with the random and free-state robustness defined similarly to the (general) robustness, but with additional restrictions~\cite{Cavalcanti2016PRA}. We recall the definitions of the general robustness 
\begin{equation*}
\begin{aligned}
\IR(\vec{M})&\coloneqq\min\{t \geq 0\;|\;\exists \vec{N}~\text{MA and}~\vec{D}\in\mathbb{JM}\ \text{s.t.}\\
&\ (M_{a|x} + t N_{a|x})/(1+t)= D_{a|x}\ \forall~a,x\},~~~~~~~~\rm{and}
\end{aligned}
\end{equation*}
\begin{equation*}
\begin{aligned}
\SR(\vec{\sigma})&\coloneqq\min\{t\geq 0\;|\; \exists\; \vec{\xi}\; \text{SA and}\; \vec{\tau}\in\LHS~\text{s.t.}\\&~(\sigma_{a|x} + t \xi_{a|x})/(1+t)=\tau_{a|x}\ \forall~ a,x\}.
\end{aligned}
\end{equation*}
For incompatibility (respectively, steering) random robustness $\IR^{\rm R}$ (respectively, $\SR^{\rm R}$), we require $N_{a|x}=\openone/|\mathcal{A}|$ (respectively, $\xi_{a|x}=\rho_{\rm{B}}/|\mathcal{A}|$) for all $a,x$, with $|\mathcal{A}|$ being the cardinality of the set of outcomes. On the other hand, the incompatibility (respectively, consistent steering) free-state robustness $\IR^{\rm{JM}}$ (respectively, $\SR^{\rm{C-LHS}}$) is defined by considering $\vec{N}\in\JM$ (respectively, $\vec{\xi}\in\LHS$ and $\sum_{a}\xi_{a|x}=\sum_a\sigma_{a|x}~\forall~x$).
The result from the previous section holds for these two types of robustness:
\begin{result}\label{R: random and free-state robustness}
Eq.~\eqref{eq:max_SR=IR}, Obs.~\ref{obs:marginal are the same}, Res.~\ref{R:inc_dist}, \ref{R:mathfilt}, \ref{R:LF1 is optimal}, \ref{R:steer_filt} still hold when $(\SR,\IR)$ is substituted with $(\SR^{\rm R},\IR^{\rm R})$ or $(\SR^{\rm C-LHS},\IR^{\rm JM})$. 
\end{result}

See Appendix H for the proof. Finally, we extend these results to the \em incompatible weight~\cite{PuseyJOPSA15} and \em steerable weight~\cite{Skrzypczyk2014PRL}, defined, respectively, as 
\begin{equation*}
\begin{aligned}
\IW(\vec{M})&\coloneqq\min\{t \geq 0\;|\;\exists \vec{N}~\text{MA and}~\vec{D}\in\mathbb{JM}\ \text{s.t.}\\
&\;M_{a|x}=t N_{a|x}+ (1-t)D_{a|x}\ \forall~a,x\},~~~~~~~~~~~~\rm{and}
\end{aligned}
\end{equation*}
\begin{equation*}
\begin{aligned}
\SW(\vec{\sigma})&\coloneqq\min\{t\geq 0\;|\; \exists\; \vec{\xi}\; \text{SA and}\; \vec{\tau}\in\LHS~\text{s.t.}\\&~\sigma_{a|x}=t \xi_{a|x}+(1-t)\tau_{a|x}\ \forall~ a,x\}.
\end{aligned}
\end{equation*}
More precisely, (in Appendix I) we show that
\begin{result}\label{R: weight}
For any MA $\vec{B}$, 
\begin{equation}\label{eq: weight_max}
\begin{aligned}
&\max_{\rho_{\rm B}} \SW\left(\rho_{\rm B}^{1/2}\vec{B}\rho_{\rm B}^{1/2}\right)=\IW (\vec{B});\\
&\SW(\rho_{\rm B}^{1/2}\vec{B}\rho_{\rm B}^{1/2})\leq \IW (\vec{B})\forall \rho_{\rm B}.
\end{aligned}
\end{equation}
Analogs of Obs.~\ref{obs:marginal are the same} and Res.\ref{R:inc_dist}, \ref{R:mathfilt}, \ref{R:LF1 is optimal}, \ref{R:steer_filt} hold for $(\SR,\IR)$ substituted with $(\SW,\IW)$.
\end{result}

Results~\ref{R: random and free-state robustness} and \ref{R: weight} both suggest that the impossibility of distilling incompatibility via local filters is general and does not depend on the specific measure of incompatibility/steerability considered, and holds, at the mathematical level, even for nonphysical maps, i.e., linear maps that are positive but not necessarily CP or trance non-increasing. Moreover, they also extend the classification of steerability, based on SEO classes and $\LFo$ filters \cite{Ku2022NC}, to the case of different measures of steerability.

%{\it Conclusions and outlook.---}
\section{Conclusions and outlook}
Resource distillation is a central problem in quantum information theory. It was explored from a theoretical and experimental perspective as well as for a wide range of physical systems and resources, e.g., entanglement~\cite{Horodecki1999,Wang2006PRL,Masanes2008PRL,GUHNE20091,Kim2011,Thinh2018PRA,Ku2021PRR}, nonlocality~\cite{Kwiat2001,Wang:20}, steering~\cite{PramanikPRA2019,Nery2020,Gupta2021PRA,Zhang2023arXiv}, and teleportation power~\cite{Horodecki2000PRA,Li2021}. Despite the importance of measurement incompatibility as a resource for several quantum information tasks \cite{PuseyJOPSA15,Heinosaari2015PRA, Haapasalo2015,Carmeli2019PRL, DesignolleNJP2019,Uola2019PRL,Paul2019PRL,Buscemi2020PRL,Uola2020PRL, Ducuara2020PRL}, the possibility of its stochastic distillation was never investigated. Here, we overcome the difficulties associated with the definition of filtered measurements and prove that measurement incompatibility can never be stochastically distilled. Our results hold for task-based incompatibility measures such as the incompatibility robustness and incompatible weight, as well as a wide range of robustness-based measures including a single measurement scenario.

Our results complete the previous investigations on the resource theory of measurement incompatibility~\cite{Heinosaari2015PRA, Carmeli2019PRL, Uola2019PRL,Paul2019PRL,Buscemi2020PRL}, showing 
a stark contrast with the case of entanglement and especially with the case of steering~\cite{Ku2022NC,Hsieh2023}.  This is rather unexpected, since measurement incompatibility and steering have been viewed as equivalent properties for years~\cite{Uola2014PRL,Quintino2014PRL,Uola2015PRL}. 
We leave to future research to explore the physics behind this difference.

Notably, our approach not only uncovers a series of no-go results for measurement incompatibility, but also solves an open problem in stochastic steering distillation.
In fact, we showed that maximum steerability is always obtainable via $\LFo$ filters, namely, local filters described by a single Kraus operator. 
This completely solves the problem of finding the optimal filter in stochastic steering distillation, since it suffices to find the optimal $\LFo$ filter, whose explicit form can be efficiently obtained via SDP as reported in Ref.~\cite{Ku2022NC}.

Many questions remain open.
First, our study focuses on incompatibility measures that can be operationally extended to filtered measurements. 
It is then interesting to investigate whether similar operational generalization can be made for different types of incompatibility~\cite{PianiJOSA2015,Ji2021arxiv,MitraPRA2022,Hsieh2022PRR,Haapasalo2021,Hsieh2023-2,Kiukas2017PRA,Ku2022PRXQ}.
For instance, whether we can stochastically distill the incompatibility of quantum instruments, i.e., the incompatibility of {\it channel assemblages} (CA) Refs.~\cite{Ji2021arxiv,MitraPRA2022}). 
A CA goes beyond an MA by considering also the output state after the measurement.

\section*{Acknowledgement}
The authors acknowledge fruitful discussions with Joonwoo Bae, Jessica Bavaresco, Shin-Liang Chen, M\'at\'e Farkas, Erkka Haapasalo, Yeong-Cherng Liang, Simon Milz, Miguel Navascues, Bartosz Regula, Yen-An Shih, Anthony Short, Paul Skrzypczyk, Ryuji Takagi, and Hao-Cheng Weng. This work is supported by the
Ministry of Science and Technology, Taiwan, (Grants No.~MOST 112-2112-M-003-020-MY3), and Higher Education Sprout Project of National Taiwan Normal
University (NTNU) and the Ministry of Education (MOE) in Taiwan, the Royal Society through Enhanced Research Expenses (on grant NFQI),  the ERC Advanced Grant (FLQuant), the Leverhulme Trust Early Career Fellowship (on grant ``Quantum complementarity: a novel resource for quantum science and technologies'' with number ECF-2024-310).

%\appendix

\begin{widetext}

\section{Appendix A: Local filters, maximum steerability, and SEO equivalence classes}\label{app:maxsup}

In this section, we recall for convenience some useful results from \cite{Ku2022NC}. These statements, in fact, were presented in the proof of the main theorems of Ref.~\cite{Ku2022NC}, but not explicitly in their statements.
The first result is the following, which can be found in the proof of Th.~2 in \cite{Ku2022NC}:
\begin{equation}\label{eq:max_rhoB=IR}
\max_{\rho_{\rm B}} \SR\left(\rho_{\rm B}^{{1/2}} \vec{B} \rho_{\rm B}^{{1/2}}\right)= \sup_{\text{$\td{\rho}_{\rm B}$: full-rank}} \SR\left(\td{\rho}_{\rm B}^{{1/2}}   \vec{B}  \td{\rho}_{\rm B}^{{1/2}}\right) = \IR(\vec{B}).
\end{equation}
It is proven by showing that from each feasible solution of one optimization problem, one can construct a feasible solution of the other optimization problem, without changing the objective function. See Eqs.~(31) and (33) in \cite{Ku2022NC} for details, and notice that ${\rm range}(\omega_{a|x})\subset {\rm range(\eta)}$ for $\eta,\vec{\omega}$ in Eq.~(31) of \cite{Ku2022NC}. In the statement of the theorem $\sup$ appeared instead of $\max$ since it is not guaranteed that $\rho_{\rm B}$ is a full-rank state, so the assemblage $\rho_{\rm B}^{{1/2}} \vec{B} \rho_{\rm B}^{{1/2}}$ may belong to a different SEO, than $\vec{B}$. Notice that, in principle, we do not even know if a full-rank solution is always achievable, since the constraint of being full-rank makes the set of variables over which we optimize non-closed and, thus, non-compact.

The second result is that
\begin{equation}
\sup_{\EE_\omega \in \LFo}\SR\left(\frac{\EE_\omega(\vec{\sigma})}{p(\omega)}\right)= \max_{\EE_\omega \in \LFo}\SR\left(\frac{\EE_\omega(\vec{\sigma})}{p(\omega)}\right)=\max_{\rho_{\rm B}} \SR(\rho_{\rm B}^{{1/2}} \vec{B} \rho_{\rm B}^{{1/2}}). 
\end{equation}
Notice that, a priori, it is not guaranteed that the supremum ($\sup$) can be achieved in the maximization over $\EE_\omega$. In our case, however, due to the upper bound given by $\max_{\rho_{\rm B}} \SR(\rho_{\rm B}^{{1/2}} \vec{B} \rho_{\rm B}^{{1/2}})$, we can explicitly construct the Kraus operator associated with $\EE_\omega$ from $\rho_{\rm B}$ and the state $\zeta_{\rm B}:=\sum_a \sigma_{a|x}$.

The third result is that 
\begin{equation}
\text{ for all } \vec{A} \ \exists\ \rho_{\rm AB} \text{ s.t. } \SR(\vec{\sigma})=\IR(\vec{A}) \text{ for } \sigma_{a|x}=\tr_{\rm A}[ A_{a|x} \otimes \openone \rho_{\rm AB} ]
\end{equation} 
Again, the proof is identical to the one presented in Th.~3 in \cite{Ku2022NC}, but uses a local filter associated with the optimal $\rho_{\rm B}$ appearing in the expression in Eq.~\eqref{eq:max_rhoB=IR}. This, together with the inequality $\SR(\vec{\sigma})\leq \IR(\vec{A})$ derived in Ref.~\cite{ShinLiang2016PRL}, proves Eq.~(4) of the main text, namely, that measurement incompatibility can be operationally defined as the maximum steerability over all bipartite states.

\section{Appendix B: Operational interpretation of a filtering measurement}

Before introducing the concept of a filtered measurement, let us first discuss the effect of measurement in quantum physics. From a physical perspective, a measurement is a procedure that, given a (quantum) system, produces an outcome with a certain probability. Hence, mathematically, a measurement maps a quantum state to a post-measurement quantum state and a distribution over measurement outcomes. Importantly, when one ignores the post-measurement state, the only relevant object is the distribution over classical outcomes.
This fact can be described by the mapping 
\begin{equation}\label{eq:meas_BR_app}
\mathcal{M} : \rho \mapsto \{ p(a)\}_a.
\end{equation}
The physics of the problem imposes several constraints on such a mapping. First, it must be that $p(a) \geq 0$ for all $a$ and $\sum_a p(a)=1$, namely, it should map states to probability distributions. Second, such a map must be linear. This comes from the fact that, if one considers a probabilistic mixture of two states, i.e., $\rho=\lambda \rho_1 + (1-\lambda) \rho_2$, with $0 < \lambda <1$, then the corresponding probability distribution must satisfy $p(a)=\lambda p_1(a) + (1-\lambda) p_2(a)$, where $p_i$ is the outcome distribution associated with the state $\rho_i$. This implies convex-linearity of the map, however, it can be extended by linearity with an argument analogous to that of \cite{Hardy2001,BarrettPRA2007}, since the objects outside of the convex set of states do not have a physical meaning.

The linearity of the map implies, by Riesz-Frechét theorem, that it can be as a scalar product on the space of linear operators (see, e.g., \cite{Heinosaari-Ziman}), namely, with the usual Born rule 
\begin{equation}\label{eq:Born_rule_app}
p(a)=\tr[M_a \rho].
\end{equation} 
Moreover, the linearity and normalization of probability, for any input state $\rho$, imply that the operators $\{M_a\}_a$ must satisfy $M_a \geq 0$, for all $a$, and $\sum_a M_a=\openone$. In other words, this is a concise derivation of the usual notion of POVM. This idea can, then, be connected with the standard measurement in textbook quantum mechanics, namely, the projective measurement via Naimark dilation theorem (see, e.g., \cite{Heinosaari-Ziman}), i.e., every POVM is implemented as a projective measurement on a larger system.

The above standard argument for the derivation of POVMs representation for quantum measurement is crucial to understand why a filtered measurement can no longer be described by a POVMs. The failure of the linearity of the mapping, which is inevitable in the case of filtering due to the presence of the post-selection on a desired outcome, implies that one must go beyond both the standard Born rule as well as the notion of POVM to describe a filtered measurement.

Before discussing the case of filters, let us first analyze the case of a deterministic transformation, i.e., a quantum channel, applied to a measurement. A quantum channel $\EE$, in the Schr\"odinger picture, is a mapping from quantum states to quantum states that is completely positive (CP) and trace-preserving (TP). The same channel can be viewed in the Heisenberg picture and it is represented by the adjoint map $\EE^\dagger$, which now maps POVMs into POVMs. In particular, $\EE^\dagger$ is CP and unital, i.e., $\EE^\dagger(\openone)=\openone$. These properties guarantee that, when applied to an MA $\vA$, gives a new MA $\vec{A}'$.

In operational terms, the channel is applied between the state preparation and the measurement, and it can be equivalently seen as a transformation of the state (Schr\"odinger picture) or a transformation of the measurement (Heisenberg picture). 
From this operational perspective, one can generalize this notion to the idea of filtering a quantum measurement, in a practical and experimentally feasible way. 
We recall that filtering a state corresponds to the application of a CP trace-non-increasing (TNI) map to a quantum state, followed by a renormalization, such that the resulting state has again trace one, namely
\begin{equation}\label{eq:filter_def}
\rho \mapsto \frac{\II(\rho)}{\tr[\II(\rho)]}.
\end{equation}
Operationally, this corresponds to post-selecting only those experimental rounds in which the filter is passed. The CPTNI map $\II$ can be thought as a part of a quantum instrument $\{\II,\II'\}$, i.e., a quantum measurement, with possible outcomes ``pass'', associated with $\II$, and ``fail'', associated with $\II'$, such that $\II+\II'$ is a CPTP map.
Without post-selection, the process remains to be deterministic due to the fact that $\II+\II'$ is a CPTP linear map. Therefore, to implement a single copy state distillation, a {\em nonlinear transformation} due to post-selection is necessary.
There is no way around it.

One can, then, apply the same reasoning for quantum channels to the case of filters from both \emph{physical} and \emph{mathematical} perspectives (see also the first paragraph in this section). 
\emph{Physically}, applying a filter between the state preparation and the measurement procedure can be equivalently seen as a part of the former or of the latter (by using Schr\"odinger and Heisenberg pictures). 
In both cases, one should take into account the filter's failure and the corresponding post-selection probability. 
Again, due to such a post-selection, the transformation of the quantum state must be nonlinear, as in Eq.~\eqref{eq:filter_def}. 
Consequently, a measurement described by a POVM cannot be mapped to another POVM in the case of postselection.
It must be mapped to a more complicated operation, which includes the filter measurement, the associated post-selection procedure, and the original measurement. 
Mathematically, applying the filter $\II_{\omega}$ to the measurement procedure $
\mathcal{M}$ in Eq.~\eqref{eq:meas_BR_app}, we obtain a new mapping, represented by the pair $(\II_{\omega}^\dagger,\{M_{a}\}_{a})$, such that the probability for the outcome $a$ with an arbitrary input state $\rho$ reads, in analogy with Eq.~\eqref{eq:meas_BR_app}, as 
\begin{equation}
\II_\omega(\mathcal{M}): \rho \mapsto \left\lbrace p_\omega(a) =  \frac{\tr[\II_\omega^\dagger(M_a) \rho]}{\tr[\II_\omega^\dagger(\openone) \rho]} \right\rbrace_a.
\end{equation}
In other words, the action of the new measurement (in terms of the probability distribution) is completely described by the following {\it unique} mapping:
\begin{equation}
(\cdot) \mapsto \tr[\II_\omega^\dagger(M_a) (\cdot)]/\tr[\II_\omega^\dagger(\openone) (\cdot)].
\end{equation}
Notice that, even if the mapping is nonlinear, as one should expect given the presence of a post-selection procedure, the definition of the new measurement mapping is {\em state-independent}. 
Hence, it is a physically well-defined (unique and state-independent) measurement procedure,  which  cannot be described by POVMs. Finally, note that this definition can be easily adapted to a set of POVMs by conditioning on a classical index set $x$ i.e., $\{M_{a|x}\}_{a,x}$.
To make a concrete example, one can consider the measurement of photon polarization after inserting a beamsplitter and performing post-selection, i.e., discarding one of the paths coming out of the beamsplitter.
This type of physical measurement, which is ubiquitous in lab practices, combines POVMs with filters. 
Hence, any discussion of filtered measurement must acknowledge this fact and generalize the notion of measurement beyond that of a POVM.

Since a filtered measurement
is no longer a POVM, we have to generalize the standard mathematical definition of incompatibility quantification for a measurement. Here, we consider operational definitions of incompatibility quantification, which allows us to study filtering measurement analytically.
First, let us recall the standard steering experiment and show that the filtered measurement can be used in a steering experiment to generate a valid SA. Alice and Bob share a bipartite system in the state $\rho_{AB}$. Alice performs a measurement on her half of the system, by selecting a measurement setting $x$ and obtaining a measurement outcome $a$. The classical information $(x,a)$ is then sent to Bob, which uses it to label the system corresponding to that round, i.e., $\sigma_{a|x}$. After repeating this procedure several times, Bob has collected an SA $\{\sigma_{a|x}\}_{a,x}$ on which he can perform state tomography (or measure a steering inequality) to decide whether the SA is steerable or not.

The case of the filtered measurement $(\II^\dagger,\{A_{a|x}\}_{a,x})$ works analogously. 
The only difference is that now Alice must inform Bob, via classical communication, either of the input-output pair $(x,a)$ or of her filter's failure. 
In the latter case, Bob can discard his system in that round. 
Now, Bob's resulting SA is simply the SA obtained if Alice and Bob were to share the bipartite state
\begin{equation}
\rho'_{AB}\coloneqq \frac{(\II \otimes{\rm id})(\rho_{AB})}{\tr[ (\II \otimes{\rm id})(\rho_{AB})]},
\end{equation}
where we denoted with ${\rm id}$ the identity channel, and Alice was to perform the original set of measurements. It is then clear that the steering experiment is well-defined also with a filtered measurement.
Importantly, as we reported in Eq.~(4) of the main text, we can further use steering experiments to evaluate and estimate incompatibility robustness. 
These allow us to use steering experiments (which are experimentally feasible and operationally clear) to quantify the effect of filters on measurements.
Finally, we remark that the operational definition of the filtered measurement also allows us to apply it to the state-discrimination task \cite{Uola2019PRL,Carmeli2019PRL,Paul2019PRL,Buscemi2020PRL} as well as for the state-exclusion task \cite{Uola2020PRL,Ducuara2020PRL}; see below for a detailed discussion.

Note that adding an additional element $\openone - \mathcal{I}_\omega^\dagger(\openone)$ can make $\mathcal{I}_\omega^\dagger(M_a)$ a complete POVM. Still, in this work, we adopt Eq.~\ref{Eq: state transform after LF} so that post-selection's effect is automatically included.
In fact, the additional POVM element $\openone-\mathcal{I}_{\omega}^\dagger(\openone)$ corresponds to the failure of the filter. This can be observed by $\mathcal{I}_{\omega}+\mathcal{I'}$ is CPTP (or, equivalently, $(\mathcal{I}^{\dagger}_{\omega}+\mathcal{I'}^{\dagger})(\openone)=\openone$ which is CP-unital map, \costa{ i.e., preserving POVMs}). In the other direction, it has been shown that measurement incompatibility cannot be distilled by considering quantum pre-processing~\cite{Buscemi2020PRL} that takes both success and failure of filters into account in the measurement transformation. Thus, merely adding an additional POVM cannot distill measurement incompatibility.

\section{Appendix C: No stochastic distillation of measurement incompatibility from state discrimination and exclusion tasks}\label{app:distill}
We explore the connection between Result 1 and state discrimination tasks. Indeed, measurement incompatibility can be operationally defined also in terms of state-discrimination tasks, i.e., for a MA $\vec{A}$ we have \cite{Uola2019PRL,Carmeli2019PRL,Paul2019PRL,Buscemi2020PRL} 
\begin{equation}\label{eq:st_discr}
1+\IR(\vec{A})=\sup_{\Ss} \frac{p_{\rm succ}^{\rm discr}(\vec{A},\Ss)}{\max_{\vec{O}\in \JM} p_{\rm succ}^{\rm discr}(\vec{O},\Ss)},
\end{equation}
where $\Ss:=\{p(a,x),\rho_{a|x}\}$ is the state ensemble involved in the state discrimination task 
with prior information \cite{Carmeli2018PRA}, and $p_{\rm succ}^{\rm discr}(\vec{A},\Ss)$ is the success 
probability associated with the MA $\vec{A}$ and the task $\Ss$. We refer the reader to 
\cite{Uola2019PRL,Carmeli2019PRL,Paul2019PRL,Buscemi2020PRL} for formal details. 
The question is how to define a generalization of this procedure when the filtering of the measurement takes place. First, we notice that the filtering operation is placed between the state preparation and the measurement. Hence, following the same steps as in the steering scenario, one can operationally interpret the filtering of the measurement as the filtering of the incoming states. 
The filtered-measurement incompatibility, then, involves the ratio between the success probability in the discrimination task with the original MA and a the optimal jointly measurable one, namely,
\begin{equation}\label{eq:filt_game}
\sup_{\Ss} \frac{p_{\rm succ}^{\rm discr}(\vec{A},\EE_\omega(\Ss))}{\max_{\vec{O}\in \JM} p_{\rm succ}^{\rm discr}(\vec{O},\EE_\omega(\Ss))},
\end{equation}
where, with a slight abuse of notation, we defined $\EE_\omega(\Ss) := \{p(a,x|\omega),\EE_\omega(\rho_{a|x})/p(\omega|a,x)\}$, and the probabilities $p(\omega|a,x)=\tr[\EE_\omega(\rho_{a|x})]$ and
\begin{equation}
p(a,x|\omega)=\frac{p(a,x,\omega)}{p(\omega)}=\frac{p(a,x)p(\omega|a,x)}{\sum_{a,x} p(\omega|a,x)p(a,x)}.
\end{equation}
Indeed, the effect of the postselection associated with the filter operation is to change the probability of appearances of the states.
From Eq.~\eqref{eq:filt_game}, it is then clear that the ratio between the two guessing probabilities cannot be increased as we are performing the same optimization, but now on a restricted set of state, i.e., the image of the filter $\EE_{\omega}$.

The same reasoning can be applied to the incompatible weight $\IW(\vec{A})$
in state exclusion tasks \cite{Uola2020PRL,Ducuara2020PRL}, namely,
\begin{equation}\label{eq:st_excl}
1-\IW(\vec{A})=\inf_{\Ss} \frac{p_{\rm succ}^{\rm excl}(\vec{A},\Ss)}{\max_{\vec{O}\in \JM} p_{\rm succ}^{\rm excl}(\vec{O},\Ss)},
\end{equation}
where $p_{\rm succ}^{\rm excl}(\vec{A},\Ss)$ refers to the success probability of such a task. One immediately sees that filter operations can only transform the state ensemble $\mathcal{S}$ into a new one, hence, they cannot decrease the infimum on the r.h.s.~of Eq.~\eqref{eq:st_excl} and, thus, they cannot increase the incompatible weight. 

Moreover, as similar argument applies also to the {\it robustness of measurement} (ROM), recently introduced in \cite{Paul2019PRL2}. The ROM is defined as the minimal amount of noise that can be added to a measurement before it becomes completely uninformative, i.e. $\ROM(M):=\min\{t| (M_a+t N_a)/(1+t)=q(a)\openone\}$. As in the previous cases, $\ROM(M)$ can be defined in terms of a discrimination task \cite{Paul2019PRL2} as
\begin{equation}
1+\ROM(M)= \sup_{\mathcal{S}} \frac{p_{\rm succ}^{\rm discr}(M,\mathcal{S})}{p_{\rm succ}^{\rm discr}(\mathcal{S})},
\end{equation}
where $\mathcal{S}:=\{p_x,\rho_x\}$ is a state ensemble, and $p_{\rm succ}^{\rm discr}(M,\mathcal{S})$ and $p_{\rm succ}^{\rm discr}(\mathcal{S})$ are the probability of guessing the correct label $x$, with and without performing the measurement $M$, respectively. Again, one can show that a filter $\EE_\omega$ on the ensemble $\mathcal{S}$ only provides a valid ensemble, so it is not able to increase $\ROM(M)$, which is already maximized over all possible $\Ss$.

Finally, we comment on the fact that this seems to be the only meaningful generalization of discrimination-based operational definitions of incompatibility, since the only the possible alternative definition has several problematic properties, as we discuss below. As such, the definition provided in Eq.~\eqref{eq:filt_game}, and the analogous definitions for the $\IW$ and $\ROM$, seem the only plausible ones. Let us first have a look at the $\ROM$ case, which is the simplest one. Now, consider as an alternative  definition the case in which the ratio is computed between the guessing probability in the filtered and unfiltered game
\begin{equation}\label{eq:filt_game2}
 \sup_{\mathcal{S}} \frac{p_{\rm succ}^{\rm discr}(M,\EE_\omega(\Ss))}{p_{\rm succ}^{\rm discr}(\mathcal{S})}.
\end{equation}
It is easy to verify that there exists very simple measurements and filters in which this value becomes arbitrarily high. Consider the case of a qubit system with the following game:  $\Ss=\{p_x, \rho_x\}_{x=0}^{N}$ with states $\rho_0=\ketbrac{0}$, and $\rho_i = \ketbrac{1}$ for $i=1,\ldots, N$, appearing with probability $p_0=\varepsilon$ and $p_i=\frac{1-\varepsilon}{N}$, for $i>0$ and some  arbitrary small $\varepsilon$. The optimal guessing probability for the case without measurement $p_{\rm succ}^{\rm discr}(\mathcal{S})$ is rather straightforward: the player guesses the state with the highest probability , i.e., $p_{\rm succ}^{\rm discr}(\mathcal{S})=\frac{1-\varepsilon}{N}$. Alternatively, one can guess at random any of the state, except the first one. It is clear that we can make such a probability arbitrary small, since $p_{\rm succ}^{\rm discr}\rightarrow 0$ for $N\rightarrow \infty$. The strategy with the measurement does not perform much better, as one can only distinguish between $\rho_0$ and all the others. The guessing probability is then $p_{\rm succ}^{\rm discr}(M,\EE_\omega(\Ss))=\varepsilon+ \frac{1-\varepsilon}{N}$. This can be obtained with the POVM $\{\ketbrac{0},\ketbrac{1}, 0, \ldots, 0\}$ or, equivalently, with $\{\ketbrac{0},\frac{1}{N}\ketbrac{1}, \ldots, \frac{1}{N}\ketbrac{1}\}$. Now, consider the game in which we implement a filter $\EE_\omega(\cdot)=\ketbrac{0} \cdot \ketbrac{0}$. The result is that the filtered game has $p_0=1$ and $p_x=0$ for $x>0$. Equivalently, one can have a slightly rotated filter, such that $p_0=1-\varepsilon'$ and $p_x=\varepsilon'/N$ for $x>0$, such that the game keeps the same number of inputs. It is then clear that, while $p_{\rm succ}^{\rm discr}(\mathcal{S})\approx 0$, $p_{\rm succ}^{\rm discr}(M,\EE_\omega(\Ss))=1$, or $\approx 1$ if we do not filter exactly with $\ketbrac{0}$. Such an operational definition of the filtered $\ROM$ is problematic for two reasons. First, one can obtain an arbitrary high value of $\ROM$, whereas, from the SDP definition of $\ROM$ one can see that it is trivially bounded by the number of outcomes of $M$ minus 1; see \cite{Paul2019PRL2}. Secondly, this increased robustness is not associated with an improved ability to discriminating the states, but rather to some ``cheating strategy'' based on manipulating the probability that they appear. 

The same argument can be applied, almost identical, to the case of $\IR$. Consider a qubit system with the MA $\vec{A}=\{ A_{a|0}\}_{a=0,1}$ defined as $A_{a|0}=\ketbrac{a}$ for $a=0,1$ and the discrimination game defined by the same state ensemble $\Ss$ as before. This is a special case of a MA, in which only one input appears, so it should not provide any advantage over jointly measurable MA. Notice that one can construct more complicated games by taking copies of it, but this fact is irrelevant for our argument. The guessing probability for the unfiltered case, which includes the $\JM$ case, is given by
\begin{equation}
\max_{\vec{O} \in \JM} p_{\rm succ}^{\rm discr}(\vec{O},\Ss)=\sum_{a=0,1} p(a) \tr[A_a \rho_a]=\varepsilon+ \frac{1-\varepsilon}{N}.
\end{equation}
Conversely, by applying the same filter as before, i.e., $\EE_\omega(\cdot)=\ketbrac{0} \cdot \ketbrac{0}$, one obtains $p_{\rm succ}^{\rm discr}(M,\EE_\omega(\Ss))=1$. Again, the ratio between the two guessing probability can grow arbitrary high, since one is constant and the other tends to $0$, in contradiction with the definition of $\IR$, for which a simple bound in terms of the number of inputs and outputs can be obtained; see \cite{Uola2015PRL}. Moreover, this increase in the ratio does not corresponds to an increasing in the ability of distinguishing state, but it comes only from ``cheating strategy'' based on the manipulation of the probability of appearance of the states. Similar considerations can be made for $\IW$, for which any value between $0$ and $1$ can be obtained, without necessarily improving the state-discrimination capabilities of the measurement.

\section{Appendix D: Proof of Observation 1}\label{Proof of marginal}
\noindent\emph{Proof.---} The main idea is that one can explicitly construct $\vec{\xi}$ and $\vec{\tau}$ from the optimal $\vec{N}$ and $\vec{D}$ appearing in the definition of $\IR$ by multiplying them by $\rho_{\rm B}^{1/2}$. Let us see this more in detail.
It is helpful to recall the definition
\begin{equation}\label{Eq:Obs1-001}
\IR(\vec{B})=\min\{t \geq 0\;|\;(B_{a|x} + t N_{a|x})/(1+t)= D_{a|x}\ \forall~a,x, \text{ for } \vec{N}~\text{MA},~\vec{D}\in\JM\ \}
\end{equation}
The central idea is that by using $\IR(\vec{B})=\max_{\rho_{\rm{B}}}\SR(\rho_{\rm{B}}^{1/2}\vec{B}\rho_{\rm{B}}^{1/2})$, one can construct a decomposition appearing in the definition of $\SR(\rho_{\rm{B}}^{1/2}\vec{B}\rho_{\rm{B}}^{1/2})$ starting from 
one in $\IR(\vec{B})$.
Let $t^*$ be the minimal $t$ achieving Eq.~\eqref{Eq:Obs1-001}; namely, $t^*=\IR(\vec{B})$. 
Also, let
$\tilde{\rho}_{\rm{B}}$ be the optimal state for $\max_{\rho_{\rm{B}}}\SR(\rho_{\rm{B}}^{1/2}\vec{B}\rho_{\rm{B}}^{1/2})$. Multiplying both sides of the $\IR(\vec{B})$ decomposition by $\tilde{\rho}_{\rm{B}}^{1/2}$, we obtain
\begin{equation}
\begin{aligned}
\tilde{\rho}_{\rm{B}}^{1/2}\vec{B}\tilde{\rho}_{\rm{B}}^{1/2} &= (1+t^*)\tilde{\rho}_{\rm{B}}^{1/2}\vec{D}\tilde{\rho}_{\rm{B}}^{1/2}- t^* \tilde{\rho}_{\rm{B}}^{1/2}\vec{N}\tilde{\rho}_{\rm{B}}^{1/2}
%\\
%&
= (1+t^*)\vec{\tau}- t^*\vec{\xi},
\end{aligned}
\end{equation}
where $\vec{\tau}=\tilde{\rho}_{\rm{B}}^{1/2}\vec{D}\tilde{\rho}_{\rm{B}}^{1/2}$ is a SA in $\LHS$ since $\vec{D}\in \JM$, and $\vec{\xi}=\tilde{\rho}_{\rm{B}}^{1/2}\vec{N}\tilde{\rho}_{\rm{B}}^{1/2}$ is a generic SA.
Note that $\vec{\tau}$ and $\vec{\xi}$ satisfy the conditions in $\SR(\tilde{\rho}_{\rm{B}}^{1/2}\vec{B}\tilde{\rho}_{\rm{B}}^{1/2})$, and we also have $t^*=\IR(\vec{B})=\SR(\tilde{\rho}_{\rm{B}}^{1/2}\vec{B}\tilde{\rho}_{\rm{B}}^{1/2})$. Consequently, $(t^*,\vec{\tau},\vec{\xi})$ provide an optimal solution for $\SR(\tilde{\rho}_{\rm{B}}^{1/2}\vec{B}\tilde{\rho}_{\rm{B}}^{1/2})$.
Finally, since $\vec{B},\vec{D},\vec{N}$ are MA, $\tilde{\rho}_{\rm{B}}^{1/2}\vec{B}\tilde{\rho}_{\rm{B}}^{1/2}$, $\vec{\tau}$ and $\vec{\xi}$ all have the same marginal state $
\tilde{\rho}_{\rm{B}}$.\hfill$\square$

\section{Appendix E: Proof of Result 2}\label{App:mathfilt}
\noindent\emph{Proof.---} The main idea is that given the optimal decomposition for the robustness on the r.h.s., i.e., either $(t,\vec{N},\vec{D})$, or $(t,\vec{\xi}, \vec{\tau})$, one can construct a valid decomposition for the l.h.s., which proves the inequality. Let us see this more in detail. 
We recall that $\SR(\vec{\sigma})=\min\{t\ge0\;|\;(\vec{\sigma}+t \vec{\xi})/(1+t)=\vec{\tau} \in \LHS, \text{ for } \vec{\xi} \text{ SA }\}$. Let $\tilde{\rho}_{\rm{B}}$ be an optimal state achieving $\IR(\vec{B})=\max_{\rho_{\rm B}}\SR({\rho}_{\rm B}^{1/2} \vec{B} {\rho}_{\rm B}^{1/2})$ and $t^*$ be the optimal parameter achieving $t^*=\SR(\tilde{\rho}_{\rm B}^{1/2} \vec{B} \tilde{\rho}_{\rm B}^{1/2})$.
By Eq. (2) in the main text, we know that the same $t^*$ appears in the decomposition of $\vec{B}$ in Eq.~\eqref{Eq:Obs1-001} as 
\begin{equation}\label{Eq:SR becomes IR}
\begin{aligned}
\vec{B}=(1+t^*)\vec{D}-t^*\vec{N},
\end{aligned}
\end{equation}
where $\vec{N}$ is a generic MA and $\vec{D}\in \JM$.
We now apply the filtering operation with an outcome $\omega$ and multiply for a generic state $\hat{\rho}_{\rm{B}}$ with the normalization $p(\omega)=\tr[\EE_\omega(\hat{\rho}_{\rm{B}})]$. We have that
\begin{equation}
\begin{aligned}
\frac{\hat{\rho}_{\rm{B}}^{1/2}\EE_{\omega}^\dagger(\vec{B})\hat{\rho}_{\rm{B}}^{1/2}}{p(\omega)}&=(1+t^*)\frac{\hat{\rho}_{\rm{B}}^{1/2}\EE_{\omega}^\dagger(\vec{D})\hat{\rho}_{\rm{B}}^{1/2}}{p(\omega)}-t^*\frac{\hat{\rho}_{\rm{B}}^{1/2}\EE_{\omega}^\dagger(\vec{N})\hat{\rho}_{\rm{B}}^{1/2}}{p(\omega)},
\end{aligned}
\label{Eq: SEO after free and max with renormalization}
\end{equation}
which is a valid decomposition for $\SR\left(\frac{\hat{\rho}_{\rm{B}}^{1/2}\EE_{\omega}^\dagger(\vec{B})\hat{\rho}_{\rm{B}}^{1/2}}{p(\omega)}\right)$.
To see this, first we note that, for {\em every} MA $\vec{M}$, we can write
\begin{align}
\tr\left[\frac{\hat{\rho}_{\rm{B}}^{1/2}\EE_{\omega}^\dagger(M_{a|x})\hat{\rho}_{\rm{B}}^{1/2}}{p(\omega)}\right] = \frac{\tr[\EE_\omega(\hat{\rho}_{\rm{B}})M_{a|x}]}{\tr[\EE_\omega(\hat{\rho}_{\rm{B}})]},
\end{align}
which is a valid conditional probability distribution.
Since $\hat{\rho}_{\rm{B}}^{1/2}\EE_{\omega}^\dagger(\vec{M})\hat{\rho}_{\rm{B}}^{1/2}\ge0$ for every $a,x$, we conclude that $\frac{\hat{\rho}_{\rm{B}}^{1/2}\EE_{\omega}^\dagger(\vec{M})\hat{\rho}_{\rm{B}}^{1/2}}{p(\omega)}$ is a valid SA for every MA $\vec{M}$.
Then all elements in Eq.~\eqref{Eq: SEO after free and max with renormalization} are valid SAs, and
$\frac{\hat{\rho}_{\rm{B}}^{1/2}\EE_{\omega}^\dagger(\vec{D})\hat{\rho}_{\rm{B}}^{1/2}}{p(\omega)}\in \LHS$ because of $\vec{D} \in \JM$.

Since the parameter $t^*$ may not be the minimum with respect to the decomposition defining $\SR\left(\frac{\hat{\rho}_{\rm{B}}^{1/2}\EE_{\omega}^\dagger(\vec{B})\hat{\rho}_{\rm{B}}^{1/2}}{p(\omega)}\right)$,  we have
\begin{equation}
\SR\left(\frac{\hat{\rho}_{\rm{B}}^{1/2}\EE_{\omega}^\dagger(\vec{B})\hat{\rho}_{\rm{B}}^{1/2}}{p(\omega)}\right)
\leq t^*=\IR(\vec{B})=\SR\left(\tilde{\rho}_{\rm{B}}^{1/2}\vec{B}\tilde{\rho}_{\rm{B}}^{1/2}\right),
\end{equation}
which is true for every $\hat{\rho}_{\rm B}$.
This concludes the proof of Eq.~\eqref{Eq:E_dag} in the main text. 

The proof of Eq.~\eqref{Eq:E} in the main text is analogous. 
For a generic state $\hat{\rho}_{\rm{B}}$,
we just apply $\EE_{\omega}$ and the normalization \mbox{$p(\omega)=\tr[\EE_\omega(\hat{\rho}_{\rm{B}})]$} to the decomposition $\vec{B} = (1+t^*)\vec{D}- t^*\vec{N}$ appearing in the definition of $\IR(\vec{B})$, where $\vec{D}\in\JM$ and $\vec{N}$ is a generic MA.
This means
\begin{equation}
\frac{\EE_\omega\left(\hat{\rho}_{\rm{B}}^{1/2}\vec{B}\hat{\rho}_{\rm{B}}^{1/2}\right)}{p(\omega)} = (1+t^*)\frac{\EE_\omega\left(\hat{\rho}_{\rm{B}}^{1/2}\vec{D}\hat{\rho}_{\rm{B}}^{1/2}\right)}{p(\omega)}-t^*\frac{\EE_\omega\left(\hat{\rho}_{\rm{B}}^{1/2}\vec{N}\hat{\rho}_{\rm{B}}^{1/2}\right)}{p(\omega)},
\end{equation} 
which is a valid decomposition for $\SR\left(\frac{\EE_{\omega}(\hat{\rho}_{\rm{B}}^{1/2}\vec{B}\hat{\rho}_{\rm{B}}^{1/2})}{p(\omega)}\right)$.
Since $\IR(\vec{B})=\SR(\tilde{\rho}_{\rm B}^{1/2} \vec{B} \tilde{\rho}_{\rm B}^{1/2})=t^*$, the result follows. \hfill$\square$

\section{Appendix F: Proof of Result 3}
\noindent\emph{Proof.---} 
Let us first recall from the main text the equations
\begin{equation}\label{eq:max_SR=IR_app}
\max_{\EE_\omega \in \LFo}\SR\left(\frac{\EE_\omega(\vec{\sigma})}{p(\omega)}\right)=\max_{\rho_{\rm B}} \SR\left(\rho_{\rm B}^{{1/2}} \vec{B} \rho_{\rm B}^{{1/2}}\right) = \IR(\vec{B}),
\end{equation}
\begin{align}
\begin{split}\label{Eq:E_dag_app}
\max_{\rho_{\rm{B}}}\SR\left(\frac{\rho_{\rm{B}}^{1/2}\EE_{\omega}^\dagger(\vec{B})\rho_{\rm{B}}^{1/2}}{p(\omega)}\right) \leq \max_{\rho_{\rm{B}}} \SR\left(\rho_{\rm{B}}^{1/2}\vec{B}\rho_{\rm{B}}^{1/2}\right),
\end{split}\\
\begin{split}\label{Eq:E_app}
\max_{\rho_{\rm{B}}}\SR\left(\frac{\EE_{\omega}(\rho_{\rm{B}}^{1/2}\vec{B}\rho_{\rm{B}}^{1/2})}{p(\omega)}\right) \leq \max_{\rho_{\rm{B}}} \SR\left(\rho_{\rm{B}}^{1/2}\vec{B}\rho_{\rm{B}}^{1/2}\right),
\end{split}
\end{align}
Fix a SA $\vec{\sigma}$, since $\LFo\subset \LF$, we have that $\max_{\EE_\omega \in \LFo}\SR\left(\frac{\EE_\omega(\vec{\sigma})}{p(\omega)}\right)\leq \max_{\EE_\omega \in \LF}\SR\left(\frac{\EE_\omega(\vec{\sigma})}{p(\omega)}\right).$
Conversely, take $\EE_\omega'\in \LF$ and take $\vec{B}$ the SEO of $\vec{\sigma}$. 
By Eqs.~\eqref{eq:max_SR=IR_app} and \eqref{Eq:E_app}, we have
\begin{align}
\SR\left(\frac{\EE_\omega'(\vec{\sigma})}{p'(\omega)}\right)\leq \max_{\rho_{\rm{B}}} \SR\left(\rho_{\rm{B}}^{1/2}\vec{B}\rho_{\rm{B}}^{1/2}\right)=\max_{\EE_\omega \in \LFo}\SR\left(\frac{\EE_\omega(\vec{\sigma})}{p(\omega)}\right)
\end{align}
 for all $\EE_\omega'$, which completes the proof.
  \hfill $\square$

\section{Appendix G: Proof of Result 4}
\noindent\emph{Proof.---}
Alice performs a measurement $\vec{A}$ on her half of $\rho_{\rm{AB}}$, generating $\vec{\sigma}$ on Bob's side. Now Alice repeats the same experiment with the filter $\{\EE_{\omega}\}_\omega$ and post-selection on an outcome $\omega$, generating the SA $\vec{\sigma'}$ for Bob. The state before her measurement is, then, given by \mbox{$\rho_{\rm{AB}}^{(\omega)}=(\EE_{\omega}^{\rm A}\otimes{\rm id}^{\rm B})(\rho_{\rm AB})/p(\omega)$,} where $\rm{id}^{\rm B}$ is the identity channel acting on ${\rm B}$ and $p(\omega)=\tr[\EE_{\omega}(\rho_{\rm{A}})]$. The corresponding SA is given by 
\begin{align}
%\begin{equation}
\label{eq:Alice_filt}
%\begin{split}
&\sigma'_{a|x}=\tfrac{1}{p(\omega)}\tr_{\rm{A}}\left[A_{a|x}\otimes \openone \left(\EE_{\omega}^{\rm{A}}\otimes{\rm id}^{\rm{B}}\right)(\rho_{\rm AB})\right]=\rho_{\rm A}^{1/2} \left[\tfrac{\EE_{\omega}^\dagger(A_{a|x})}{p(\omega)}\right]^T \rho_{\rm A}^{1/2}
=   \left[\rho_{\rm B}^{1/2}\tfrac{\EE_{\omega}^\dagger(A_{a|x})}{p(\omega)} \rho_{\rm B}^{1/2}\right]^T,
\end{align}
%\end{split}
%\end{equation}
where we used that $\rho_{\rm AB}$ is a pure state, in the second equality to write the assemblage (this results follows directly from the Schmidt decomposition of the pure state $\rho_{\rm AB}=\sum_{ij} \lambda_i \lambda_j \ketbra{ii}{jj}$; see \cite{Uola2015PRL} for more details) and the in the third to say that $\rho_{\rm A}=\rho_{\rm B}$, and the transposition is taken with respect to the basis where $\rho_{\rm B}$ is diagonal. An analogous calculation shows that $\sigma_{a|x}=[\rho_{\rm B}^{1/2}A_{a|x} \rho_{\rm B}^{1/2}]^T$. We identify Eq.~\eqref{eq:Alice_filt}, up to a global transposition of the assemblage which does not change its steerability, with the l.h.s.~of Eq.~\eqref{Eq:E_dag_app}, while the r.h.s.~is $\SR(\vec{\sigma})$, by Eq.~\eqref{eq:max_SR=IR_app}, Result~3, and the assumption that the initial steerability of the assemblage is maximal over $\LF$ filters.
\hfill$\square$

\section{Appendix H: Robustness-based measures and proof of Result 5}\label{App:free and random robustness}
Here, we extend the result from \cite{Ku2022NC}, namely, $\IR(\vec{B})=\max_{\rho_{\rm{B}}}\SR(\rho_{\rm{B}}^{1/2} \vec{B} \rho_{\rm{B}}^{1/2})$, to other measures of robustness, such as the random and free-state consistent robustness, both in the incompatibility and steering scenarios.

Let us recall the definition of {\em random incompatibility robustness}~\cite{Cavalcanti2016PRA}:
\begin{equation}
\begin{aligned}
    \IR^{\rm R}(\vec{B})=\min\left\{
    %&
    t\ge0\;\middle|\;\frac{B_{a|x}+t\openone/|\mathcal{A}|}{1+t}=\sum_{\lambda}p(a|x,\lambda)G_{\lambda},%\\
    %&
    G_{\lambda}\geq 0, \sum_\lambda G_{\lambda}=\openone\right\},
\end{aligned}
\end{equation}
where $|\mathcal{A}|=n_a$ is the number of the outcomes $a$.
On the other hand, the {\em random steering robustness} (see the definition in Ref.~\cite{Cavalcanti2016PRA}) of a SA $\vec{\sigma}$ with reduced state \mbox{$\sum_a\sigma_{a|x}=\rho_{\rm B}$} can be expressed as
\begin{equation}
\begin{aligned}
    \SR^{\rm R}(\vec{\sigma})=\min\left\{
    %&
    t\ge0\;\middle|\;\frac{\sigma_{a|x}+t\rho_{\rm B}/|\mathcal{A}|}{1+t}=\sum_{\lambda}p(\lambda) p(a|x,\lambda)\rho_{\lambda},%\\
    %&
    \rho_{\lambda}\geq 0, \tr[\rho_\lambda]=1\right\}.
\end{aligned}
\end{equation}

Let us also recall the definition of free-state robustness measures. The {\em JM incompatibility robustness} is given by (see, e.g., \cite{Cavalcanti2016PRA})
\begin{equation}
\begin{aligned}
    \IR^{\rm{JM}}%&
    (\vec{B})=\min\left\{ t\ge0\;\middle|\;\frac{B_{a|x}+t\sum_{\lambda}p'(a|x,\lambda)G'_{\lambda}}{1+t}=\sum_{\lambda}p(a|x,\lambda)G_{\lambda};%\\ &
    \text{ with }
    G_{\lambda}, G'_{\lambda}\geq 0\ \forall \lambda, \sum_\lambda G_{\lambda}=\sum_\lambda G'_{\lambda}=\openone\right\}.
\end{aligned}
\end{equation}
Similarly, the {\em consistent LHS steering robustness} (see the definition in~\cite{Cavalcanti2016PRA}) of a SA $\vec{\sigma}$ with reduced state \mbox{$\sum_a\sigma_{a|x}=\rho_{\rm B}$} can be expressed as
\begin{equation}
\begin{aligned}
    \SR^{\rm{C-LHS}}%&
    (\vec{\sigma})=\min\left\{ t\ge0\;\middle|\;\frac{\sigma_{a|x}+t\sum_{\lambda}p'(\lambda)p'(a|x,\lambda)\rho'_{\lambda}}{1+t}=\sum_{\lambda}p(\lambda)p(a|x,\lambda)\rho_{\lambda};%\\ &
    \text{  $\rho_{\lambda}, \rho'_{\lambda}$: states s.t. $\sum_\lambda p(\lambda)\rho_\lambda=\rho_{\rm B}$}\right\}.
\end{aligned}
\end{equation}

\subsection{Lemmas relevant to Result 5}
Before proceeding to the proof of Result 5, we need some intermediate results. More precisely, we need to prove that 
\begin{lemma}\label{lemma:free-stateR}
The maximal consistent LHS steering robustness within an SEO class is the same as the JM incompatibility robustness of the SEO; namely
\begin{equation}
    \IR^{\rm{JM}}(\vec{B})=\max_{\rho_{\rm{B}}}\SR^{\rm{C-LHS}}\left(\rho_{\rm{B}}^{1/2}\vec{B}\rho_{\rm{B}}^{1/2}\right).
\label{result2}
\end{equation}
Moreover, every full-rank state is an optimal $\rho_{\rm{B}}$.
\end{lemma}

\noindent\emph{Proof.---}
Let $t^* = \IR^{\rm{JM}}(\vec{B})$ with the decomposition $\vec{B} = (1+t^*)\vec{D}- t^*\vec{D}'$ appearing in the definition of $\IR^{\rm JM}(\vec{B})$; that is, both $\vec{D},\vec{D}'$ are in $\JM$.
Then, for every $\rho_{\rm B}$, we have 
\begin{align}
\rho_{\rm B}^{1/2}\vec{B}\rho_{\rm B}^{1/2} = (1+t^*)\rho_{\rm B}^{1/2}\vec{D}\rho_{\rm B}^{1/2}- t^*\rho_{\rm B}^{1/2}\vec{D}'\rho_{\rm B}^{1/2},
\end{align}
which is a valid decomposition for $\SR^{\rm{C-LHS}}\left(\rho_{\rm{B}}^{1/2}\vec{B}\rho_{\rm{B}}^{1/2}\right)$.
This shows that $\max_{\rho_{\rm{B}}}\SR^{\rm{C-LHS}}\left(\rho_{\rm{B}}^{1/2}\vec{B}\rho_{\rm{B}}^{1/2}\right)\le\IR^{\rm{JM}}(\vec{B})$.

To prove the tightness of this inequality, let us consider an arbitrary {\em full-rank} state $\hat{\rho}_{\rm B}$.
Let us write $\hat{t} = \SR^{\rm{C-LHS}}\left(\hat{\rho}_{\rm B}^{1/2}\vec{B}\hat{\rho}_{\rm B}^{1/2}\right)$ with the following decomposition:
\begin{align}
\hat{\rho}_{\rm B}^{1/2}\vec{B}\hat{\rho}_{\rm B}^{1/2} = \left(1 + \hat{t}\right)\vec{\xi} - \hat{t}\vec{\tau},
\end{align}
where, by the definition of $\SR^{\rm C- LHS}$, both $\vec{\xi},\vec{\tau}$ are in $\LHS$, and both of them have the same reduced state $\hat{\rho}_{\rm B}$ that is full-rank.
By applying the map $\hat{\rho}_{\rm B}^{-1/2}(\cdot)\hat{\rho}_{\rm B}^{-1/2}$, we can write
\begin{align}
\vec{B} = \left(1 + \hat{t}\right)\vec{D} - \hat{t}\vec{N},
\end{align}
where $\vec{D},\vec{N}$ are the SEOs of $\vec{\xi},\vec{\tau}$, meaning that both of them are in $\JM$.
This is therefore a valid decomposition for $\IR^{\rm{JM}}(\vec{B})$.
Consequently, we conclude that 
\begin{align}
\IR^{\rm{JM}}(\vec{B})\le\hat{t} = \SR^{\rm{C-LHS}}\left(\hat{\rho}_{\rm B}^{1/2}\vec{B}\hat{\rho}_{\rm B}^{1/2}\right)\le\max_{\rho_{\rm{B}}}\SR^{\rm{C-LHS}}\left(\rho_{\rm{B}}^{1/2}\vec{B}\rho_{\rm{B}}^{1/2}\right),
\end{align}
which completes the proof.

 \hfill$\square$\\

 It is worth mentioning that our proof also shows that {\em every full-rank} state can be used to construct the optimal solution.
 This makes our proof also constructive, since we know how to explicitly choose an optimal $\rho_{\rm B}$.
 By using the same proof, a similar result also holds for the random robustness:
\begin{lemma}
The maximal random steering robustness within an SEO class is the same as the random incompatibility robustness of the SEO; namely
\begin{equation}
    \IR^{\rm R}(\vec{B})=\max_{\rho_{\rm{B}}}\SR^{\rm R}\left(\rho_{\rm{B}}^{1/2}\vec{B} \rho_{\rm{B}}^{1/2}\right).
\label{result3}
\end{equation}
Moreover, every full-rank state is an optimal $\rho_{\rm{B}}$.
\end{lemma}

\noindent\emph{Proof.---} One can use exactly the same proof of Lemma~\ref{lemma:free-stateR} to show this result.
More precisely, the argument goes through by replacing the free-state robustness with random robustness.
\hfill$\square$\\

\subsection{Proof of Result 5}
To prove Result 5, we need to show that all the statements we made for the pair $(\IR,\SR)$ can be reformulated for the pairs $(\SR^{\rm R},\IR^{\rm R})$ and $(\SR^{\rm C-LHS},\IR^{\rm JM})$. The above lemmas show that 
Eq.~\eqref{eq:max_SR=IR} in the main text holds with the substitution of $(\IR,\SR)$ with $(\SR^{\rm R},\IR^{\rm R})$ and $(\SR^{\rm C-LHS},\IR^{\rm JM})$. The same holds for the analogous of Obs. 1, by the definitions of $\SR^{\rm C-LHS}$ and $\SR^{\rm R}$. Similarly, one can prove the analogous of Res. 2, 3, 4, since the original proof is based on Eq.~\eqref{eq:max_SR=IR} in the main text and Obs. 1. Regarding Res. 1, one can proceed by showing that also Eq.~\eqref{eq:IR_sup_SR} in the main text can be transformed into a version
\begin{equation}
\widetilde{\IR}^{Y}(\vec{A}) := \max_{\rho_{\rm AB}}\ \SR^{X}(\vec{\sigma}(\rho_{\rm AB})),
\end{equation}
for the pairs $(X,Y)=({\rm C-LHS},{\rm JM})$ or $({\rm R},{\rm R})$. and repeat the same argument that leads to Res. 1.\hfill $\square$

\section{Appendix I: Weight-based measure and proof of Result 6}\label{App:weight-based measure}
In this section, we extend our Obs. 1 and Res. 2, 3, 4 from robustness-based measures to the weight-based measures. 
We now recall their definition. The  {\em incompatible weight} \cite{PuseyJOPSA15}  is defined as
\begin{equation}
\begin{aligned}\label{Eq: def of IW}
\IW(\vec{B})=\min\Big\lbrace& t\ge0 \;\Big|\;B_{a|x}=t N_{a|x}+(1-t)O_{a|x}~\forall a,x,\ \vec{N} \text{~~is any MA, }\vec{O}\in\JM\Big\rbrace.
\end{aligned}
\end{equation}
Similarly, the {\em steerable weight} \cite{Skrzypczyk2014PRL} can be expressed as
\begin{equation}
\begin{aligned}\label{eq:def_SW}
\SW(\vec{\sigma})=\min\Big\lbrace& t\ge0 \;\Big|\;\sigma_{a|x}=t \xi_{a|x}+(1-t)\tau_{a|x}~\forall a,x,\ \vec{\xi}  \text{~~is any SA},\vec{\tau}\in\LHS\Big\rbrace.
\end{aligned}
\end{equation}

Before we show our statements in this section, we first derive the primal SDP of the $\IW$ by reformulating the optimization problem as
$N_{a|x}=\frac{B_{a|x}-(1-t)O_{a|x}}{t}\geq 0$.
If we define $(1-t)=s$, we have
$B_{a|x}-s\sum_{\lambda}D(a|x,\lambda)G_{\lambda}\geq 0 $.
Here, we use the definition of the joint measureability $O_{a|x}=\sum_{\lambda}D(a|x,\lambda)G_{\lambda}$ with $G_{\lambda}$ being a POVM satisfying $\sum_{\lambda}G_{\lambda}=\openone$.
Finally, by defining $\tilde{G}_{\lambda}=sG_{\lambda}$, we  arrive at  the primal form of the SDP
\begin{equation}
    \begin{aligned}
    \text{Given}~~&\vec{B}\\
    \text{Find}~~&\max s=1-\IW(\vec{B}) %\tr\left(\sum_{a,x}\omega_{a|x}B_{a|x}\right):=1+\IR(\vec{B}) 
    \\
    \text{s.t.}~~ &B_{a|x}\geq\sum_{\lambda}D(a|x,\lambda)\tilde{G}_{\lambda},\\
    &\tilde{G}_{\lambda}\geq 0~~\text{and}~~\sum_{\lambda}\tilde{G}_{\lambda}=s\openone.
    \end{aligned}
    \label{eq:primal_IW}
\end{equation}
Here, the last constraint is from $\sum_{\lambda}\tilde{G}_{\lambda}=s\sum_{\lambda}G_{\lambda}=s\openone$.
The Lagrangian associated with the primal SDP is
\begin{equation}
\begin{aligned}
    L&=s+\sum_{a,x}\tr\left[\omega_{a|x}\left(B_{a|x}-\sum_{\lambda}D(a|x,\lambda)\tilde{G}_{\lambda}\right)\right]%\\&
    +\tr\left[Z\left(s\openone-\sum_{\lambda}\tilde{G}_{\lambda}\right)\right]+\sum_{\lambda}\tr\left[Y_{\lambda}\tilde{G}_{\lambda}\right]\\&
    =s(1+\tr\left[Z\right])+\sum_{a,x}\tr\left[\omega_{a|x}B_{a|x}\right]%\\&
    +\sum_{\lambda}\tr\left[\tilde{G}_{\lambda}\left(-\sum_{a,x}D(a|x,\lambda)\omega_{a|x}-Z+Y_{\lambda}\right)\right].
    \end{aligned}
\end{equation}
Here,  $Z,\omega_{a|x}$, and $Y_{\lambda}$ are the Hermitian dual variables.
One can directly check that
\begin{align}
1-\IW(\vec{B}) = \max_{\substack{s\ge0,\\\{\tilde{G}_{\lambda}\ge0\}}}\min_{\substack{Z = Z^\dagger,\\\{\omega_{a|x}\ge0\},\\\{Y_{\lambda}\ge0\}}}L\le\min_{\substack{Z = Z^\dagger,\\\{\omega_{a|x}\ge0\},\\\{Y_{\lambda}\ge0\}}}\max_{\substack{s\ge0,\\\{\tilde{G}_{\lambda}\ge0\}}}L,
\end{align}
where the upper bound is the dual SDP.
To write it down, note that
the Lagrangian  is independent of the primal variable if 
$\tr\left[Z\right]=-1$ and $Y_{\lambda}= \sum_{a,x}D(a|x,\lambda)\omega_{a|x}+Z$.
Together with the criteria $\omega_{a|x}\ge0,Y_\lambda\geq 0$, we obtain
the dual SDP of the $\IW$ as follows, where we define a new variable $X\coloneqq - Z$,
\begin{equation}
    \begin{aligned}
    \text{Given}~~&\vec{B}\\
    \text{Find}~~&
    \min \sum_{a,x}\tr\left[\omega_{a|x}B_{a|x}\right]=1-\IW(\vec{B}) 
   % \max s=1-\IR(\vec{B}) %\tr\left(\sum_{a,x}\omega_{a|x}B_{a|x}\right):=1+\IR(\vec{B}) 
    \\
    \text{s.t.}~~ &\sum_{a,x}D(a|x,\lambda)\omega_{a|x}\ge X,\\  
&\tr\left[X\right]=1,
%%%~~\text{and}~~
\omega_{a|x}\geq 0~\forall a,x.
    \end{aligned}
    \label{eq:dual_IW_2}
\end{equation}
We note that in general $X$ is not a quantum state because the first constraint in Eq.~\eqref{eq:dual_IW_2} does not guarantee $X\geq 0$.

On the other hand, the dual SDP of the steerable weight $\SW$ is~\cite{Skrzypczyk2014PRL}
\begin{equation}
    \begin{aligned}
    \text{Given}~~&\vec{\sigma}\\
    \text{Find}~~&
    \min \sum_{a,x}\tr\left[F_{a|x}
\sigma_{a|x}   \right]=1-\SW(\vec{\sigma}) 
   % \max s=1-\IR(\vec{B}) %\tr\left(\sum_{a,x}\omega_{a|x}B_{a|x}\right):=1+\IR(\vec{B}) 
    \\
    \text{s.t.}~~ &\sum_{a,x}D(a|x,\lambda)F_{a|x}\geq \openone,\\  
&F_{a|x}\geq 0~~ \forall~a,x.
    \end{aligned}
    \label{Eq:dual SDP of SW}
\end{equation}

\subsection{Lemmas relevant to Result 6}

To show the main result of this subsection, we first show the following lemma:
\begin{lemma}\label{L:IW>=SW}
    Given a MA $\vec{B}$, the weight-based measures satisfy
    \begin{equation}
    \SW\left(\rho_{\rm B}^{1/2}\vec{B}\rho_{\rm B}^{1/2}\right) \leq \IW(\vec{B})\; \text{ for all } \rho_{\rm B}. 
    \end{equation}
\end{lemma}
\noindent\emph{Proof.---}
We apply the positive map $\rho_{\rm B}^{1/2}(\cdot)\rho_{\rm B}^{1/2}$, where $\rho_{\rm B}$ is an arbitrary quantum state, to the elements of the decomposition appearing in Eq.~\eqref{Eq: def of IW}. Then we define 
$\vec{\xi}=\rho_{\rm B}^{1/2}\vec{N}\rho_{\rm B}^{1/2}$ and  $\vec{\tau}=\rho_{\rm B}^{1/2}\vec{O}\rho_{\rm B}^{1/2}$. The decomposition in $\IW$, then, becomes a valid decomposition for $\SW(\rho_{\rm B}^{1/2}\vec{B}\rho_{\rm B}^{1/2})$, since $\vec{O}\in\JM$ implies that $\rho_{\rm B}^{1/2}\vec{O}\rho_{\rm B}^{1/2}\in\LHS$, which proves the claim.
\hfill$\square$\\

We now show our main result in this subsection:
\begin{lemma}\label{obs:IW=SW}
The maximal steerable weight in the SEO class is equivalent to the incompatible weight of the SEO; i.e.,
\begin{equation}\label{Eq:maxrhoBSW=IW}
\max_{\rho_{\rm{B}}}\SW\left(\rho_{\rm{B}}^{1/2}\vec{B}\rho_{\rm{B}}^{1/2}\right)=\IW(\vec{B}),
\end{equation}
where the optimal $\rho_{\rm{B}}$ can be obtained by solving the dual SDP of $\IW(\vec{B})$.
\end{lemma}
\noindent\emph{Proof.---} By Lemma~\ref{L:IW>=SW}, it is enough to prove that any feasible solution of the problem on the r.h.s.~of Eq.~\eqref{Eq:maxrhoBSW=IW} allows us to construct a feasible solution of the problem on the l.h.s.~with a lower objective function. In other words, we want to prove the opposite inequality with respect to Lemma~\ref{L:IW>=SW}. Let us start with the feasible solutions, $X$ and $\vec{\omega}$, of $\IW(\vec{B})$ defined in Eq.~\eqref{eq:dual_IW_2}. Since the operator $X$ is hermitian, it can always be decomposed as $X=X^+-X^-$ satisfying $X^\pm\geq 0$ and $X^+X^-=X^-X^+=0$. Because of $\tr(X)=1$, we have $\tr(X^+)\geq1$. We then define the following variables:
\begin{align}
\eta_{\rm{B}}=\frac{X^+}{\tr(X^+)},\quad\vec{\sigma}=\eta_{\rm{B}}^{1/2}\vec{B}\eta_{\rm{B}}^{1/2},\quad\vec{F}=\eta_{\rm{B}}^{-1/2}\vec{\omega}\eta_{\rm{B}}^{-1/2},
\end{align}
where the inverse is defined after a projection on the range of $\eta_{\rm{B}}$, which is also the range of $X^+$.
Let $\Pi_{X^+}$ be the projector onto $X^+$'s range. Then we have $\eta_{\rm{B}}^{1/2}\eta_{\rm{B}}^{-1/2}=\eta_{\rm{B}}^{-1/2}\eta_{\rm{B}}^{1/2}=\Pi_{X^+}$.
The objective function of Eq.~\eqref{eq:dual_IW_2} can be re-written by

\begin{align}\label{Eq:IWSWcomp011}
\sum_{a,x}\tr\left(B_{a|x}\omega_{a|x}\right)&=\sum_{a,x}\tr\left[B_{a|x}\omega_{a|x}\left(\Pi_{X^+} + (\id-\Pi_{X^+})\right)\right]\ge\sum_{a,x}\tr\left(B_{a|x}\omega_{a|x}\Pi_{X^+}\right)=\sum_{a,x}\tr\left[B_{a|x}\left(\Pi_{X^+} + (\id-\Pi_{X^+})\right)\omega_{a|x}\Pi_{X^+}\right]\nonumber\\
&\ge\sum_{a,x}\tr\left(B_{a|x}\Pi_{X^+}\omega_{a|x}\Pi_{X^+}\right)=\sum_{a,x}\tr\left(\eta_{\rm{B}}^{1/2}{B}_{a|x}\eta_{\rm{B}}^{1/2}\eta_{\rm{B}}^{-1/2}\omega_{a|x}\eta_{\rm{B}}^{-1/2}\right)=\sum_{a,x}\tr\left(\sigma_{a|x}F_{a|x}\right),
\end{align}
which is the objective function of $1-\max_{\rho_{\rm{B}}}\SW\left(\rho_{\rm{B}}^{1/2}\vec{B}\rho_{\rm{B}}^{1/2}\right)$. See Eq.~\eqref{Eq:dual SDP of SW} for comparison.

We now show that the constraints in Eq.~\eqref{Eq:dual SDP of SW} are satisfied.
Applying the CP map $\eta_{\rm{B}}^{-1/2}(\cdot)\eta_{\rm{B}}^{-1/2}$ on $\sum_{a,x}D(a|x,\lambda)\omega_{a|x}\ge X$, which is the first constraint in Eq.~\eqref{eq:dual_IW_2}, we obtain
\begin{align}
\sum_{a,x}D(a|x,\lambda)F_{a|x}\ge\eta_{\rm{B}}^{-1/2}X\eta_{\rm{B}}^{-1/2}=\eta_{\rm{B}}^{-1/2}X_+\eta_{\rm{B}}^{-1/2}
%%%\eta_{\rm{B}}^{-\frac{1}{2}}(X^+-X^-)\eta_{\rm{B}}^{-\frac{1}{2}}=\eta_{\rm{B}}^{-\frac{1}{2}}(X^+)\eta_{\rm{B}}^{-\frac{1}{2}}
=\tr[X^+]\openone\geq\openone.
\end{align}
Notice that, without loss of generality, we can restrict the space in Eq.~\eqref{Eq:dual SDP of SW} to be the range of $X^+$, since by construction it includes the ranges of both $\sigma_{a|x}$ and $F_{a|x}$ for all $a,x$.
Consequently, $\vec{F}$ is a feasible solution to Eq.~\eqref{Eq:dual SDP of SW}, meaning that
\begin{align}
\sum_{a,x}\tr\left(\sigma_{a|x}F_{a|x}\right)\ge 1-\SW\left(\eta_{\rm{B}}^{1/2}\vec{B}\eta_{\rm{B}}^{1/2}\right)\ge1-\max_{\rho_{\rm{B}}}\SW\left(\rho_{\rm{B}}^{1/2}\vec{B}\rho_{\rm{B}}^{1/2}\right).
\end{align}
Together with Eq.~\eqref{Eq:IWSWcomp011}, we conclude that
\begin{align}
\sum_{a,x}\tr\left(B_{a|x}\omega_{a|x}\right)\ge1-\max_{\rho_{\rm{B}}}\SW\left(\rho_{\rm{B}}^{1/2}\vec{B}\rho_{\rm{B}}^{1/2}\right)
\end{align}
for {\em every} feasible solutions $X$ and $\vec{\omega}$ of $\IW(\vec{B})$'s dual SDP in Eq.~\eqref{eq:dual_IW_2}.
In other words, by minimizing over all feasible solutions, we obtain
%%%\begin{align}
$1-\IW(\vec{B})\ge1-\max_{\rho_{\rm{B}}}\SW\left(\rho_{\rm{B}}^{1/2}\vec{B}\rho_{\rm{B}}^{1/2}\right)$; namely,
%%%\end{align}
\begin{align}
\IW(\vec{B})\le\max_{\rho_{\rm{B}}}\SW\left(\rho_{\rm{B}}^{1/2}\vec{B}\rho_{\rm{B}}^{1/2}\right),
\end{align}
which,using Lemma~\ref{L:IW>=SW}, concludes the proof.
This also means that optimal solutions to $\IW(\vec{B})$'s dual SDP in Eq.~\eqref{eq:dual_IW_2} can help us to construct optimal state $\rho_{\rm B}$.
\hfill$\square$\\

\subsection{Proof of Result 6}
The first part of the proof, i.e., Eq.~\eqref{eq: weight_max} in the main text, comes from the above lemmas. The impossibility of distillability of $\IW$ has been proven in Supp. C~\ref{app:distill}. To conclude, we need to show that Obs. 1  and Res. 2, 3, and 4 hold for $(\IR,\SR)$ substituted with $(\IW,\SW)$. 
For Obs. 1, it is enough to show that from any solution of $\IW(\vec{B})$ we can construct an assemblage  $\vec{\sigma}=\eta_{\rm{B}}^{1/2}\vec{B}\eta_{\rm{B}}^{1/2}$ as in Lemma~\ref{obs:IW=SW} above. The corresponding solution for $\SW(\vec{\sigma})$ has the SA $\vec{\xi}$ and $\vec{\tau}$ appearing in Eq.~\eqref{eq:def_SW} with the same reduced state as $\vec{\sigma}$.
For Res. 2, one can proceed similarly to the proof in Supp. E~\ref{App:mathfilt} and construct a convex decomposition for the SA: $\rho_{\rm B}^{\frac{1}{2}} \EE_\omega^\dagger(\vec{B}) \rho_{\rm B}^{\frac{1}{2}}/p(\omega)$, necessary to compute its $\SW$, from the convex decomposition of $\vec{B}$. One proceeds similarly as in  Supp. E~\ref{App:mathfilt} also for the SA: $\EE_\omega(\rho_{\rm B}^{1/2} \vec{B} \rho_{\rm B}^{1/2})/p(\omega)$.
For Res. 3, it is enough to use Res. 2 together with Lemma.~\ref{obs:IW=SW} and the fact that $\max_{\rho_{\rm{B}}}\SW\left(\rho_{\rm{B}}^{1/2}\vec{B}\rho_{\rm{B}}^{1/2}\right)=\max_{\EE \in \LFo}\SW\left(\frac{\EE_\omega(\vec{\sigma})}{p(\omega)}\right)$, since the $\LFo$ filter operation can be directly computed from the optimal $\rho_{\rm B}$ as in the case of $(\SR,\IR)$ (see also Ref.~\cite{Ku2022NC}). From this, it follows directly also Res. 4 by repeating the same construction as in the case of $(\SR,\IR)$. \hfill $\square$

\end{widetext}

%\bibliography{Ref.bib}

\end{document}